# Dynamics and Interactions of Binaries and Neutron Stars in Globular Clusters

*Steinn Sigurdsson*[1] *and E.S. Phinney*

130–33, California Institute of Technology, Pasadena, Ca 91125
steinn@mail.ast.cam.ac.uk, esp@tapir.caltech.edu
[1]current address: Institute of Astronomy, Madingley Road, Cambridge CB3 0HA, UK.

# Dynamics and Interactions of Binaries and Neutron Stars in Globular Clusters

*Steinn Sigurdsson*[1] *and E.S. Phinney*
130–33, California Institute of Technology, Pasadena, Ca 91125
steinn@mail.ast.cam.ac.uk, esp@tapir.caltech.edu
[1]current address: Institute of Astronomy, Madingley Road, Cambridge CB3 0HA, UK.

*ABSTRACT*

We model the dynamics of test binaries in isotropic, multi-mass models of galactic globular clusters. The evolution of binary orbits through the cluster potentials is modeled, including second order diffusion terms, and probabilities for close encounters with field stars are calculated. We carry out Monte Carlo simulations of the effects of the binary–single star encounters on the binary population and distribution in the cluster, and estimate the collision rate for different stellar populations in globular clusters with different structural parameters. In particular, we consider the rate for neutron stars and massive white dwarfs to undergo mass transfer due to collisional encounters, and hence estimate the population of observable millisecond pulsars in different clusters formed by binary–single star encounters. Assuming a Salpeter IMF, for low concentration clusters the core encounter rate is dominated by turnoff mass main–sequence stars and medium mass white dwarfs. For high concentration, high density clusters the encounter probabilities are increasingly dominated by neutron stars and heavy white dwarfs. Hence we predict a smaller ratio of blue stragglers and cataclysmic variables to pulsars in high concentration clusters. The total number of millisecond pulsars, and the ratio of single to binary pulsars, is broadly consistent with the observed population, suggesting the binary–single star encounters contribute significantly to the pulsar formation rate in globular clusters, for the whole range of globular cluster types. The number of millisecond pulsars and the ratio of pulsars in different globular clusters is best explained by a total binary fraction comparable to that of the galaxy, and a modest number of primordial neutron stars in the globular clusters.

# 1. Introduction

## *1.1. Globular Clusters*

Globular clusters provide an excellent test for models of self–gravitating systems conjured by dynamicists. Containing large enough a number of bodies to be well described by a continuous phase space distribution, yet few enough that two–body relaxation is important in their dynamical evolution; they are the grandest computational challenge of collisional astrophysical systems.

As the technology improves, it has become possible to model the effects of stellar collisions and binary encounters in globular clusters, both in terms of the impact on the evolution of the globular cluster, and the production of "exotic" stellar species, such as blue stragglers (BSs), cataclysmic variables (CVs) and millisecond pulsars (MSPs) (King 1962, 1966, Michie 1963, Verbunt & Meylan 1987). Following the discovery of a number of X–ray sources (Katz 1975, Hertz & Grindlay 1983, Hertz & Wood 1985, Lewin & Joss 1981, Predehl, P., Hasinger, G. & Verbunt, F. 1991), and, later, millisecond pulsars (Hamilton *et al.* 1985, Lyne *et al.* 1987, Phinney & Kulkarni 1994) in Galactic globular clusters, it has become apparent that there is more to globular cluster dynamics than earlier models have allowed for.

It has long been appreciated that the static models of globular clusters developed (King 1962, 1966, Michie 1963, Michie & Bodenheimer 1963) are an approximation, and that the gravothermal instability (Lynden–Bell & Wood 1968, Katz 1978) would inevitably lead to cluster mass segregation and core–collapse (Makino & Hut 1991, Spitzer 1987, Binney & Tremaine 1987, and references therein). Extensive simulations of globular cluster evolution, initially with single mass models, later with more sophisticated multi–mass models, strongly suggests that core–collapse is inevitable, and would occur typically in a few half–mass relaxation times (Spitzer & Har 1971b, Spitzer & Thuan 1972, Cohn 1979, 1980, Murphy & Cohn 1988). Observed cluster profiles suggest that a remarkable number of clusters are very near core–collapse (Djorgovski & Hut 1992). Unless we live at a special time in the evolution of the Galaxy, this would appear, *a priori*, to be unlikely. The problem is obviated if core–collapse can be postponed or reversed; to do so requires an energy source to heat the core. Amongst the many possible energy sources for slowing core–collapse, binary interactions appear to be the most robust (Elson *et al.* 1987, Goodman 1989, Goodman & Hut 1989).

Approximating cluster stars as point masses must fail in the core–collapse limit. As the core density approaches infinity, the finite size of stars becomes important, and dissipative effects start to dominate stellar encounters. The formation of hard binaries through three–body interactions and tidal capture, and the subsequent interactions of the binaries can halt and then reverse the collapse of the cluster core (Goodman 1987, Statler *et al.* 1987), albeit at very high central densities. If the core is dominated by single degenerate stars, binary formation through three–body interactions dominates. Evidently binaries formed during core collapse cannot be part of the mechanism postponing core–collapse during the earlier evolutionary phases of the cluster. A possible energy source for postponing core–collapse is the presence of primordial binaries (Hut 1983c, Goodman & Hut 1989, Hut *et al.* 1992).

The binary abundance in the Galaxy and halo has been estimated to be no less than 20%, and possibly as high as 50% (Abt 1983, 1987, Latham 1989). Our understanding of stellar formation is not sufficiently developed to state that the initial cluster binary abundance must be similar, but we cannot say with any confidence that there **cannot** have been a primordial binary population in the Galactic globular clusters. An early search for spectroscopic giant binaries found no evidence for the existence of a binary population in the globular clusters (Gunn and Griffin 1979), but subsequent observations have found a number of binaries (Pryor *et al.* 1985, 1989, Mateo *et al.* 1990, Murphy *et al.* 1991, Bolte 1991, 1992, Yan & Mateo 1994), and current observations are consistent with a primordial binary abundance of 10% or more (Pryor 1989). We will argue below that selection effects and binary dynamics conspire to decrease the number of observable binaries, and that the primordial binary abundance in globular clusters may have been as high as the observed Galactic abundance.

In addition to the intrinsic interest in the effect of cluster binaries on the structure and evolution of the cluster, the presence of a substantial primordial binary population may in large part account for the detection of a large number of X–ray sources and pulsars in Galactic globular clusters.

## 1.2. Compact objects in globular clusters

There are twelve classic Low Mass X–ray Binaries (LMXBs) in the Galactic globular clusters (Lewin & Joss 1981, Predehl, P., Hasinger, G. & Verbunt, F. 1991 van Paradijs 1993). Comparing the hundred odd LMXBs in the Galaxy with the number of LMXBs per unit mass in the globular cluster system ($\sim 10/10^{7.5}\,\mathrm{M}_\odot$, compared to $\sim 100/10^{11.5}\,\mathrm{M}_\odot$ in the Galaxy), the number of cluster LMXBs per unit mass appears quite excessive. The launch of *Rosat* has led to the discovery of more cluster X–ray sources, and there are indications that a number of faint, soft X–ray sources are also present in the clusters (Charles 1989, Grindlay 1994). LMXBs are canonically thought to be progenitors of millisecond pulsars, and it was soon realised that the abundance of LMXBs might indicate a similar excess of pulsars in clusters (Alpar *et al.* 1982). The first cluster MSP was soon found (Hamilton *et al.* 1985, Lyne *et al.* 1987), and intense searches have now revealed a large number of MSPs and binary MSPs in clusters (Phinney & Kulkarni 1994, van den Heuvel 1991). In the Galaxy, a comparison of the inferred birthrates of LMXBs and MSPs suggested that there was an excess of MSPs relative to the LMXBs (Narayan *et al.* 1990, Lorimer *et al.* 1993). In the globular clusters, this excess is also present, and possibly worse (Bailyn & Grindlay 1990, Kulkarni *et al.* 1990). It seems clear that another class of MSP progenitors may exist; in the Galaxy this second channel for MSP formation may be through massive Be stars (Verbunt 1990, Johnston *et al.* 1992), a channel that is not available in the globular cluster population; a different mechanism for MSP formation must be invoked for the cluster pulsars.

That there are neutron stars in globular clusters is evident from the observation of MSPs. As there is no star formation taking place currently in Galactic globular clusters, the neutron stars must be primordial, or, possibly, recently formed by accretion induced collapse (AIC) of heavy white dwarfs (Michel 1987, Grindlay & Bailyn 1988). In either instance mass transfer must have taken place recently: if the MSP ancestors are dead, primordial neutron stars, they must accrete to spin up; if the MSP ancestors are heavy white dwarfs, they must accrete to pass over the Chandrasekhar limit (Nomoto & Kondo 1991). LMXBs are believed to be accreting neutron stars, and hence are good candidates for being the progenitors of at least some millisecond pulsars. The hard tidal capture binaries thought to form during core–collapse are an obvious source of LMXBs; if the star captured is a neutron star, it will be captured in an orbit likely to lead to mass transfer; if the captor is a main–sequence star near turnoff, or a (sub)giant, as is relatively probable in a globular cluster, stellar evolution will also drive mass transfer onto the neutron stars, again leading to a LMXB (Verbunt 1990). Indeed, six of twelve classic cluster LMXBs are in clusters thought to have undergone core–collapse (Trager *et al.* 1993, Grindlay 1993); four of the remaining LMXBs are in very dense clusters, which may have gone through core–collapse.

LMXBs have long inferred lifetimes (but see Tavani 1991); and consequently low inferred birthrates, they are also readily detectable in even the most distant cluster. Although many LMXBs are transient, and may not have been observed in their on–state, this does not affect the birthrate argument, as they presumably also do not accrete in the off–state. There are now thirty two reported MSPs in the Galactic globular clusters (Taylor *et al.* 1993). MSPs are hard to detect (Johnston & Kulkarni 1991, Johnston *et al.* 1991); allowing for selection effects, the true number of pulsars in globular clusters is expected to be much higher. A number of MSPs have been found in relatively low–density clusters (Kulkarni *et al.* 1990b), in proportions far in excess of those expected from the two body tidal capture scenario (Fruchter & Goss 1990, Johnston *et al.* 1991, Phinney & Kulkarni 1994).

A possible solution of the MSP birthrate problem is presented by a population of primordial binaries. A binary in a background of stars will undergo occasional close encounters with the field stars. The possible outcome of such encounters can be usefully classified by the total center–of–mass energy of the three–body system (Heggie 1975, Hills 1975a,b, Hut & Bahcall 1983), parametrized by the ratio of the relative velocity at infinity and a critical velocity, $v_\infty/v_c$.

$$v_c = \sqrt{\frac{G m_T}{a}\frac{\mu_{12}}{m_3}}, \tag{1.1}$$

where, $\mu_{12}$ is the binary reduced mass, $m_3$ is the mass of the field star encountering the binary, and $m_T$ is the total mass of the three stars. Crudely, for $v_\infty/v_c \gtrsim 1$, energy is transferred to the binary, for $v_\infty/v_c \lesssim 1$ energy is transferred to the field star (see Sigurdsson & Phinney 1993). In globular clusters, $v_\infty \sim 10$ km s$^{-1}$, and binaries with semi–major axis $\lesssim 10\,AU$ are effectively hard for stellar masses $\sim 0.5 - 1.0\,\mathrm{M}_\odot$.

An encounter may lead to a change of state in the binary: the original binary may emerge intact with different eccentricity and semi–major axis; one of the members of the binary may be exchanged, leaving the field star as a member of the new binary; there may be tidal encounter or collision between a pair of stars, or the binary may be ionized if $v_\infty/v_c \geq 1$, that is the system may become unbound, leaving three single stars. For encounters with $v_\infty/v_c < 1$, the encounter may be resonant (Hut & Bahcall 1983). Collisions are relatively more probable during resonant encounters. Cross–sections for various encounter scenarios have been calculated and tabulated extensively (Heggie 1975, Hills 1975a,b, Hills & Fullerton 1980, Fullerton & Hills 1982, Hut & Bahcall 1983, Hut 1983a, Hut & Inagaki 1985, Mikkola 1983, 1984a,b, McMillan 1986, Rappaport *et al.* 1990, Leonard 1989, Leonard & Fahlman 1991, Sigurdsson & Phinney 1993, Davies *et al.* 1994). While collision cross-sections are of general interest, and provide a measure of binary interactions in globular clusters, they fail in two ways to provide a consistent measure of the true physics: the encounters are not drawn from a population representative of the true local population of field stars, and, the history of the binary in the globular cluster is not tracked, in particular the recoil due to encounters and the consequent change in binary distribution is not followed (Phinney & Sigurdsson 1991).

## 1.3. *Stellar collisions in globular clusters*

Of particular interest in globular clusters are encounters leading to stellar collisions. A neutron star (or white dwarf) colliding with a main–sequence star or a (sub)giant is likely to disrupt the star leaving a thick disk around the degenerate (Finzi 1978, Krolik *et al.* 1984, Benz *et al.* 1987, 1989, 1990, Ruffert & Müller 1990, Davies *et al.* 1991, Davies *et al.* 1994, Rasio & Shapiro 1991, Lai *et al.* 1993, Goodman & Hernquist 1991, Sigurdsson & Hernquist 1992). Before the disk is disrupted, a substantial amount of matter may accrete onto the degenerate at a very high rate. It is possible that enough matter may be accreted to spin the neutron star up to a MSP (Type I encounters, as classified by Kochanek [1991]). In the case of a white dwarf, it is not clear if accretion past the Chandrasekhar limit is possible, and if so, whether a neutron star is formed at that point (Verbunt *et al.* 1989). Main–sequence star mergers through this channel may, at least in part, account for the "blue stragglers" observed in globular clusters (Leonard 1989, Leonard & Fahlman 1991, Livio 1993, Sigurdsson *et al.* 1994). The cross–section for three–body encounters is more weakly dependent on the core density than the two–body tidal capture scenario, and is consistent with observation of MSPs in globular clusters of different density (Fruchter & Goss 1990, Johnston *et al.* 1991, Phinney & Kulkarni 1994). In addition, the time evolving distribution of binary parameters affects the encounter rate, as we attempt to elucidate in part in this paper (see also Hut, McMillan & Romani 1992).

As a cluster collapses, the core density increases, and the binary encounter rate increases. For the stellar mass–ratios expected in cluster binaries, encounters tend to provide positive feedback in the initial stages, leading to an increasing encounter rate, as some binaries are softened, absorbing energy from the cluster. Later in the evolution of the cluster, binaries undergo exchanges which both widen the binary and increase gravitational focusing, increasing the encounter (and exchange) rate. As the cluster approaches core–collapse this feedback is negated, as binaries start to be disrupted by collisions, or become hard enough that encounters lead to the binaries being ejected on wide, eccentric orbits about the cluster core (Phinney and Sigurdsson 1991). At all stages up to the disruption of the cluster, a residue of binaries will still be making excursions to the cluster core for the first time, having resided in the cluster halo, with relaxation times of the order of the Hubble time. The energy input from the binaries will slow down the cluster collapse, postponing core–collapse. As the remaining binaries harden, the encounter rate decreases, and each encounter becomes more likely to eject the binary from the core, reducing the energy input from the binaries. Adjusting to the reduced energy input, the core contracts and core–collapse continues.

In this paper, we consider the explicit time evolution of a population of test binaries in a fixed cluster background. A comparison between the collision rate in different cluster models is obtained, as is the expected energy input to the cluster due to binary encounters. Previous estimates of the energy input from binaries have been made from analytic approximations and averaged cross–section (Murphy *et al.* 1990, Gao *et al.* 1991, Hut, McMillan & Romani 1992) and did not allow for the feedback as binaries undergo rate enhancing encounters, nor the actual probability distribution of encounter parameters, as detailed below. Though our treatment of binary–single star encounters is exact, it should be pointed out that we do not allow for binary–binary encounters. To include a binary population would increase the number of mass groups ($N_m$) to $O(N_m^2)$, and the computation effort by a corresponding factor. If the number of mass groups is large, the implicit population average represented by the distribution function becomes a poor representation of the true distribution, and the dynamics are dominated by statistically unique events not necessarily representative of any real physical system. The equilibrium distribution of the binaries is not known *a priori*, and for dense clusters is most likely a function of the binary semi–major axis. We hope that our calculations will provide information enabling binary–binary interactions to be included in a consistent manner in future simulations. We note that stable hierarchical trinaries formed in binary–binary encounters may be an important collision channel in moderate density clusters, and should be accounted for in a complete cluster simulation.

## 2. Cluster models

The binaries were evolved in a fixed background cluster model, defined by the density profile $n(\mathbf{r})$ of its stars of mass $m_\alpha$. The model can be adapted for any density profile and associated gravitational potential and velocity distribution functions. In this paper we consider isotropic multi-mass Michie–King models only. Future calculations will consider evolution in a more general model, including time varying density distributions.

In general, we considered cluster models defined by one particle distribution functions $f_\alpha(\mathbf{x}, \mathbf{v}, m_\alpha)$, for a discrete set of mass groups, $m_\alpha$, $\alpha = 1, \cdots, N_m$, with corresponding local number densities, $n_\alpha$, and mass densities, $\rho_\alpha$. Initially we considered analytic, static models for the distribution function; later we hope to develop more consistent models with a time varying distribution function. In particular, in this paper, we only consider multi–mass Michie–King models,

$$f_\alpha(\varepsilon) = \frac{n_{0_\alpha}}{(2\pi\sigma_\alpha^2)^{3/2}}\left[e^{\varepsilon/\sigma_\alpha^2} - 1\right], \tag{2.1}$$

where $\sigma_\alpha^2$ is the core dispersion of mass group $\alpha$, and $\varepsilon = -\Phi - \frac{1}{2}m_\alpha\,\mathbf{v}^2$ is the energy of the particle in the cluster center–of–mass frame. $n_{0_\alpha}$ are normalising constants to be determined later. Since $\rho_\alpha = m_\alpha n_\alpha$, we define number and mass total densities, $n(\mathbf{x}) = \sum n_\alpha(\mathbf{x})$, $\rho(\mathbf{x}) = \sum \rho_\alpha(\mathbf{x})$, and mean core mass $\bar{m}_c$ and mean core dispersion, $\bar{\sigma}^2$, scaling to the individual mass groups,

$$\begin{aligned} \bar{m}_c &= \rho_0^{-1}\sum_{\alpha=1}^{N_m} m_\alpha\rho_\alpha(0) \\ m_\alpha\sigma_\alpha^2 &= \bar{m}_c\bar{\sigma}^2, \end{aligned} \tag{2.2}$$

where $\rho_0 = \rho(0)$. Hence we define a scale radius $r_0$, analogous to the King radius in single mass King models,

$$r_0 = \sqrt{\frac{9\bar{\sigma}^2}{4\pi G\rho_0}}. \tag{2.3}$$

Each mass group is then scaled independently with scale radius

$$r_{0_\alpha} = \sqrt{\frac{\bar{m}_c}{m_\alpha}}r_0. \tag{2.4}$$

The models reduce to single mass King models for $N_m = 1$. As a check of the consistency of our code we verified that the models created did indeed reduce exactly to the corresponding King models in that limit. We have also used the solution to generate complete N–body realizations of multi-mass Michie–King models, with one particle of the appropriate mass, position and velocity for each star. These N–body models were generated for simulations of mass loss in young globular clusters due to stellar evolution, neutron star retention in globular clusters and the effects of tidal shocking on globular clusters. As part of our simulations we have run the models as isolated systems for up to 300 dynamical times and verified the distribution is virial and stable (Sigurdsson & Hernquist in preparation).

We define a new potential, $\Psi(r) = \Phi(r_t) - \Phi(r)$, where $\rho(r \geq r_t) = 0$. Note reversal of sign, $\nabla\Psi = -4\pi G\rho$. Following the King model analogy, we define a dimensionless parameter, $W_0 = \Psi(0)/\bar{\sigma}^2$, and concentration $c = \log_{10}(r_t/r_0)$. The total mass of the cluster, $M$, is then given by

$$-\nabla\Psi(r_t) = \frac{GM}{r_t^2}. \tag{2.5}$$

To obtain realistic cluster models, we need to solve for $n_{0_\alpha}$ and $r_t$, given $r_0$, $W_0$, $n(0) = n_0$, $m_\alpha$, $M$ and $\bar{\sigma}^2$. In practice, not all the quantities are independent, and some are scale invariant. We pick scale $n_0 = 1$, $r_0 = 1$, and $\bar{\sigma}^2 = 1$, and choose $W_0$ as an independent parameter determining $r_t/r_0$, for the given initial mass function. Our choice of $n_0$ and $\bar{\sigma}^2$ then uniquely determines $M$ and $r_0$ in physical units.

To solve for $n_{0_\alpha}$ and $r_t$, we must decide a cluster mass function and the relative abundance of each mass group. The current set of calculations assumes a Salpeter IMF,

$$\frac{dN_*}{dm_*} \propto m_*^{-1-x_*}, \tag{2.6}$$

with canonical value $x_* = 1.35$, though we also used $x_* = 1.0, 1.5$ in some of the simulations described here. In the current calculation, the zero age main–sequence number fraction of each mass group was calculated for a mass range, typically $0.1\mathrm{M}_\odot \leq m_* \leq 15\mathrm{M}_\odot$. A turnoff mass was selected, $0.8\mathrm{M}_\odot$; stars below the turnoff mass were assumed not to evolve significantly on the time scale of the simulation, stars above the turnoff were assumed to have evolved completely before the start of the simulation. Later simulations will allow for explicit stellar evolution during the dynamical evolution. Other initial mass functions, such as Meylan's broken IMF (Meylan 1988), were also used to generate models to compare with those used.

Stars with mass above the turnoff mass, but below some critical mass, $m_{wd}\,(= 4.7\mathrm{M}_\odot)$, were assumed to have evolved to white dwarfs of mass $0.58 + 0.22 \times (m_* - 1.0)\mathrm{M}_\odot$, while stars above mass $m_{wd}$ but below some critical mass, $m_I\,(= 8.0\mathrm{M}_\odot)$, were assumed to disrupt completely; stars with mass greater then $m_I$, but less then $m_{bh}$, were assumed to become neutron stars of mass approximately $1.4\mathrm{M}_\odot$ (Chernoff and Weinberg 1990). The mass function was truncated at $m_{bh}$, and no black holes were assumed to be present. For the purposes of the current set of simulations, all evolved stars were assumed to have been retained in the cluster; in practice some fraction is expected to be ejected, in particular a substantial fraction of the neutron stars may be ejected. Bins for mass groups use the same binning used in Phinney (1992), and the mass fraction and number fraction, $\eta_\alpha$, in each bin were calculated by integrating the evolved initial mass function.

To solve for $n_{0_\alpha}$, we followed the method of Da Costa and Freeman (1976, see also Gunn and Griffin 1979). A trial solution $n_{0_\alpha} = \eta_\alpha$ was used, the cluster model integrated, the actual number fraction of each mass group in the cluster,

$$N_\alpha = 4\pi \int_0^{r_t} n(r)\, r^2\, dr, \tag{2.7}$$

was calculated and compared with $\eta_\alpha$; a new solution

$$n_{0_\alpha} = n_{0_\alpha} \times \left(\frac{\eta_\alpha}{N_\alpha}\right)^j, \tag{2.8}$$

$j = 1$, was then substituted, and the integration iterated.

To integrate the cluster model, we integrated for $\Psi$ in radial coordinates $\xi = \log_{10}\,(1 + r^2/r_0^2)$, with implicit scaling $n(0) = 1$, $\bar{\sigma}^2 = 1$. Substituting into equation 2.1 using equation 2.3, we obtain

$$\begin{aligned}\frac{d}{d\xi}\left[\left(\frac{2(e^\xi - 1)^{3/2}}{e^\xi}\right)\frac{d}{d\xi}\right]\Psi(\xi) = -\sum_\alpha \frac{e^\xi(e^\xi - 1)^{1/2}}{2} 9 n_{0_\alpha} \frac{1}{\tilde{\rho}_\alpha} \\ \times \left[e^{\frac{\Psi(\xi)}{\sigma_\alpha^2}}\,\mathrm{erf}\Big(\sqrt{\frac{\Psi(\xi)}{\sigma_\alpha^2}}\Big) - \sqrt{\frac{4\Psi(\xi)}{\pi\sigma_\alpha^2}}\Big(1 + \frac{2\Psi(\xi)}{3\sigma_\alpha^2}\Big)\right],\end{aligned} \tag{2.9}$$

where $\tilde{\rho}_\alpha$ are normalising constants,

$$\tilde{\rho}_\alpha = \frac{\bar{m}_c}{m_\alpha}\left[\mathrm{erf}\Big(\sqrt{\frac{\Psi(0)}{\sigma_\alpha^2}}\Big) e^{\frac{\Psi(0)}{\sigma_\alpha^2}} - \sqrt{\frac{4\Psi(0)}{\pi\sigma_\alpha^2}}\Big(1 + \frac{2\Psi(0)}{3\sigma_\alpha^2}\Big)\right]. \tag{2.10}$$

Equation 2.7 was integrated using a simple leapfrog integrator with boundary conditions $\Psi(0) = W_0$, and $\nabla\Psi(0) = 0$ (in practice the integration was started at finite $r$ with $\nabla\Psi(\delta r) \propto \delta r$). $\xi$ was incremented in variable steps up to $r = r_t$, defined by $\Psi(r = r_t) = 0$. Stepsize was proportional to $r$ for $r < 1$, and constant in $\xi$ for $r > 1$, providing the highest density of steps near $r = 0$ and near the core boundary. Typically the converged model required $O(10^2)$ integration steps, although the intermediate integrations often required more integration steps. The model was considered to have converged when $\max|1 - \eta_\alpha/N_\alpha| < \delta\,(= 10^{-3})$.

Cluster parameters were not found to be sensitive to $\delta$ for small $\delta$, nor was there significant variation in the cluster parameters when the integration step was reduced by an order of magnitude, indicating that a robust solution had been found. After convergence, $n(r)$, $\Psi(r)$ and $\nabla\Psi(r)$ were saved, as were each of $n_\alpha(r)$. A cubic spline fit was also made to each of the quantities above (using standard IMSL or NR spline fitting routines), and the breakpoints and coefficients for each fit were saved for future use. In addition the mass density, $\rho_\alpha(r)$ and the projected surface density $\Sigma_\alpha(r)$ were calculated and saved. Density profiles obtained were compared with previous published calculations (Gunn and Griffin 1979, Da Costa and Freeman 1976, Meylan 1988) and were found to be in good agreement, providing an additional test of the models. The half mass radius $r_h$ was also calculated, as was the dispersion profile, $\sigma^2_\alpha(r)$. Given a luminosity function, a surface brightness profile can be constructed, here we assume the light profile is dominated by turnoff mass stars and (sub)giants.

A set of models was constructed by choosing $W_0$ and an initial mass function. The density profiles and concentrations were calculated, and comparison with real cluster profiles was made, selecting appropriate $n_0$ and $\bar{\sigma}^2$, in an attempt to reproduce observed core radii, core dispersion and cluster concentrations. The choice of IMF was dictated by recent results from pulsar acceleration limits (Phinney 1991), in particular for models of 47Tuc we chose not to use the models fit by Meylan (1989), in anticipation that they underestimate the number of neutron stars in the core. Measurements of pulsar period derivatives in the near future should severely constrain the cluster mass function. The mass groups produced by the evolved IMF are described in Table 1. A number of cluster models were used, the ones discussed in this paper are described in Table 2. The high mean core mass of the shallow IMF models used, flattens the turnoff mass star core profile, forcing a choice of larger $W_0$ then typically used in single mass models, in order to reproduce the observed concentrations of the globular clusters of interest. The high $W_0$ inferred has interesting implications for the high velocity interlopers seen in some clusters (Meylan *et al.* 1991).

# 3. Dynamics

## 3.1. Initial Conditions

An ensemble of binaries was evolved in model globular clusters, for a fixed length of time, $T$, using a cluster of DECstation 3100s and, more recently, IBM RS 6000s. We evolved binaries of mass $m_b$ $(= m_1+m_2)$, eccentricity $e$, semi–major axis, $a$, in a fixed background potential, $\Psi(r)$, as calculated in the previous section. The binary masses were drawn independently and at random from the initial mass function, with a proviso that we may require $m_i > m_{i_{min}}$, and specify the stellar type. The initial eccentricity was selected from a distribution, $P(e) = 2e$, except that we required any binary containing a (sub)giant to have initial $e = 0$, irrespective of binary period or evolutionary stage of the giant. Observations suggest that Population II binaries in the galaxy with orbital periods less then 10 days also have $e = 0$ (Abt 1983), but as close encounters will perturb the eccentricity away from zero, we decided not to impose that condition. It should be noted that this produces some apparently abnormal binaries in some of the models, for example one run consisting of a white dwarf–main sequence star binary with an eccentricity of 0.98! While such a system would not be expected to occur through normal stellar evolution, it is a possible product of an exchange encounter, and thus such a system may plausibly exist in an evolved cluster. The algorithm setting the eccentricity of binaries containing giants to zero, was triggered by some heavy white dwarf and neutron star containing binaries, for which the minimum primary mass was greater than the giant mass, resulting in a number of hard, zero–eccentricity binaries containing degenerates, mimicking a population of both unperturbed and exchanged binaries. The semi–major axis was selected from a uniform $\log(a)$ distribution, $a_{min} \leq a \leq a_{max}$. After selecting $m_{1,2}$, the stellar type was determined (main–sequence, giant/subgiant, white dwarf or neutron star) by comparing a random number with the fractional abundance of each stellar type in that mass–group. The stellar type was coded with an integer flag, and the stellar radius, $R_*$, was calculated. We assumed

$$R_* = \beta\Big(\frac{m_\alpha}{\mathrm{M}_\odot}\Big)^\alpha \mathrm{R}_\odot, \tag{3.1}$$

with $\alpha = 1.0 = \beta$ for main–sequence stars, $\alpha = -1/3$, $\beta = 0.0162$ for white dwarfs (Shapiro and Teukolsky 1983), and $\beta = 0$ for neutron stars. In reality $\alpha$ is somewhat less than unity for main–sequence stars, but for these calculations the approximation $\alpha = 1$ is adequate.

As we did not allow for explicit stellar evolution, (sub)giants were assumed to occur with constant probability for any star in the turnoff mass group ($0.63\mathrm{M}_\odot \leq m \leq 0.8\mathrm{M}_\odot$). The total probability of a star in that mass group being a (sub)giant was assumed to be 0.095 (fraction of cluster age (sub)giant of that mass lasts), with a distribution of stellar radius such that $t(R_* > R) \propto R^{-3/2}$. Assuming a giant lifetime of $4.7 \times 10^7$ years for $R_* = 10\mathrm{R}_\odot$ (Fahlman *et al.* 1985), for a power–law IMF we get a probability distribution for the fraction of stars in the turnoff mass group with radius $> R$, $f_g$,

$$f_g(R_* > R) = \frac{4.7\times10^7}{7\times10^{10}} \frac{0.8^{-1-x_*}}{\frac{1}{x_*}\big((0.63^{-x_*}) - (0.8^{-x_*})\big)} \left(\frac{R_*}{10\mathrm{R}_\odot}\right)^{-3/2}. \tag{3.2}$$

After calculating the stellar radius, if $2(R_{*_1} + R_{*_2}) \leq a(1-e)$, we required $a = a + f_c(R_{*_1} + R_{*_2})/(1-e)$, in order to avoid immediate merger of the stars. Unless stated otherwise, $f_c = 2$. In the absence of a better understanding of binary formation we assumed the binary members may be picked independently from the evolved IMF (Tout 1991), but we note that in practice binary masses may be correlated; in particular, mass transfer in the protostar phase, and during giant evolution for binaries containing evolved stars, may bias the mass function.

The binary was placed in the cluster at radius, $\mathbf{r}$, selected from the density distribution, $\rho_\alpha(r)$, of one of the mass groups, with velocity, $\mathbf{v}$, picked from the local dispersion for that mass group. Both the initial positions and velocities were assumed to be isotropic in the cluster center–of–mass frame. The initial distribution was deliberately chosen not to be the relaxed equilibrium distribution of stars mass $m_b$, in order to permit the binaries to relax naturally, particularly in the case of the lower concentration models. The initial distribution in the more concentrated, "evolved" models was more concentrated, but somewhat less concentrated than a population of point masses of comparable mass would be. As the simulations reveal, the binary population never settles down to the distribution expected for point mass stars of the same mass, as recoil induced by encounters ejects hard binaries from the core of the cluster.

To pick the velocity, the peak of the velocity distribution was estimated,

$$\max_v f_\alpha(v) = f_\alpha(v_m),$$

for the mass group chosen, with

$$\begin{aligned} v'^2 &= 2\left(1 - e^{-\frac{\Psi(r)}{2\sigma_\alpha^2}}\right) \\ v_m^2 &= 2\left(1 - e^{\frac{v'^2}{2} - \frac{\Psi(r)}{2\sigma_\alpha^2}}\right), \end{aligned} \tag{3.3}$$

and a velocity was chosen from the distribution by Monte–Carlo acceptance–rejection, scaled down by the relative mass of the binary.

The binaries were evolved in the cluster center–of–mass frame according to

$$\ddot{\mathbf{r}} = \nabla\Psi(r) + \mathbf{a_{dyf}} + \mathbf{a_{kick}}, \tag{3.4}$$

where $\nabla\Psi(r)$ is the potential gradient due to the mass interior to $r$, $\mathbf{a_{dyf}}$ is the dynamical friction experienced by the binary, and $\mathbf{a_{kick}}$ is the effective acceleration due to scattering by individual stars in the cluster. The kicks are time averaged, and a large kick may be due to a single close encounter or the cumulative effect of many smaller perturbations from distant stars. For binaries, close encounters will strongly perturb the binary orbit, and significant energy transfer may take place, sufficient to eject one or both of the binary and encountered star from the cluster. We calculate the effects of such close encounters separately and explicitly.

To calculate $\mathbf{a_{dyf}}$ and $\mathbf{a_{kick}}$, we first calculated the diffusion coefficients, $D(\Delta(v_i))$, $D(\Delta v_i \Delta v_j)$ (Binney and Tremaine 1987). We consider a local orthonormal basis, $\{\hat{\xi}, \hat{\zeta}, \hat{\eta}\}$, relative to the binary, defined by the binary's position, $\mathbf{r}$, and velocity, $\mathbf{v}$, in the cluster center–of–mass frame, with unit vectors

$$\begin{aligned} \hat{\xi} &= \frac{\mathbf{v}}{v} \\ \hat{\zeta} &= \frac{\mathbf{r} \times \mathbf{v}}{|\mathbf{r} \times \mathbf{v}|} \\ \hat{\eta} &= \frac{(\mathbf{r} \cdot \mathbf{v})\mathbf{v} - v^2 \mathbf{r}}{v\sqrt{r^2 v^2 - (\mathbf{r} \cdot \mathbf{v})^2}}, \end{aligned} \tag{3.5}$$

$\mathbf{r} = \{x, y, z\}$, $\mathbf{v} = \{v_x, v_y, v_z\}$, then force components along the $\hat{\xi}$ direction are parallel to the binary's direction of motion, and $\hat{\zeta}$, $\hat{\eta}$ define two equivalent (by symmetry) components perpendicular to the direction of motion. Hence we have three independent diffusion coefficients, $D(\Delta v_{\|})$, $D(\Delta v_{\|}^2)$ and $D(\Delta v_{\perp}^2)$.

Defining

$$\mu = \frac{4\pi}{9} \frac{GM}{r_0 \bar{\sigma}^2}, \tag{3.6}$$

we find

$$\begin{aligned} \mathbf{a_{dyf}} &= D(\Delta v_{\|}) \\ &= \sum_\alpha \frac{81}{4\pi} \frac{(m_b + m_\alpha)}{M} \mu \ln(\Lambda) \frac{1}{v^2} \frac{1}{\tilde{\rho}_\alpha} \\ &\times \left( \mathrm{erf}\left(\frac{v}{\sqrt{2\sigma_\alpha^2}}\right) e^{\frac{\Psi}{\sigma_\alpha^2}} - \frac{v}{\sigma_\alpha} e^{\frac{\Psi}{\sigma_\alpha^2} - \frac{v^2}{2\sigma_\alpha^2}} - \frac{1}{3}\left(\frac{v}{\sigma_\alpha}\right)^3 \right), \end{aligned} \tag{3.7}$$

where $\ln\Lambda \sim 10$ is the Coulomb logarithm. Some care must be maintained in evaluating $\mathbf{a_{dyf}}$ near turning points in the orbit, as finite precision in evaluating the integral can produce sign errors in the dynamical friction, especially at evaluations of intermediate steps in high order integration schemes. This can lead to spurious systematic expansion of the orbit in the cluster. This can be a serious problem for nearly radial trajectories, integrated with maximal timesteps, as is necessary to complete the calculation in finite CPU time. To deal with the underflow, the dynamical friction can either be set to zero, or the sign reversed. As the magnitude of the dynamical friction is always small when a sign error may occur, either method is adequate. The other two components are given by

$$
\begin{aligned}
D(\Delta v_{\parallel}^2) = & \sum_{\alpha} \frac{27}{\sqrt{2\pi}} \frac{m_\alpha}{M} \mu \ln(\Lambda) \frac{1}{v} \frac{1}{\tilde{\rho}_\alpha} \\
& \times \left( \frac{3}{2} \sqrt{2\pi} \frac{\sigma_\alpha^2}{v^2} \operatorname{erf}(\frac{v}{\sqrt{2\sigma_\alpha^2}}) e^{\frac{\Psi}{\sigma_\alpha^2}} - 3 \frac{\sigma_\alpha}{v} e^{\frac{\Psi}{\sigma_\alpha^2} - \frac{v^2}{2\sigma_\alpha^2}} \right. \\
& \left. - \frac{v}{\sigma_\alpha} \left( \frac{\Psi}{\sigma_\alpha^2} - \frac{3}{10} \frac{v^2}{\sigma_\alpha^2} + 1 \right) \right) \\
D(\Delta v_{\perp}^2) = & \sum_{\alpha} \frac{81}{\sqrt{2\pi}} \frac{m_\alpha}{M} \mu \ln(\Lambda) \frac{1}{v} \frac{1}{\tilde{\rho}_\alpha} \\
& \times \left( \sqrt{\frac{\pi}{2}} \left(1 - \frac{\sigma_\alpha^2}{v^2}\right) \operatorname{erf}(\frac{v}{\sqrt{2\sigma_\alpha^2}}) e^{\frac{\Psi}{\sigma_\alpha^2}} + \frac{\sigma_\alpha}{v} e^{\frac{\Psi}{\sigma_\alpha^2} - \frac{v^2}{2\sigma_\alpha^2}} \right. \\
& \left. - \frac{2}{3} \frac{v}{\sigma_\alpha} \left( \frac{\Psi}{\sigma_\alpha^2} - \frac{1}{10} \frac{v^2}{\sigma_\alpha^2} + 1 \right) \right).
\end{aligned}
\tag{3.8}
$$

We model $\mathbf{a_{kick}}$ by random fluctuations in velocity, $\Delta \mathbf{v}$,

$$
\mathbf{a_{kick}} = \frac{\Delta \mathbf{v}}{\Delta t}, \tag{3.9}
$$

with

$$
\begin{aligned}
\Delta v_{\parallel}^2 &= \varsigma_i^2 (D(\Delta v_{\parallel}^2) \Delta t) \\
\Delta v_{\perp}^2 &= \varsigma_i^2 (D(\Delta v_{\perp}^2) \Delta t),
\end{aligned}
\tag{3.10}
$$

where $\varsigma_i$ is a random number of mean 0, standard deviation 1, chosen here from a normal distribution. Assuming isotropy, in our coordinate system, this becomes

$$
\begin{aligned}
\Delta v_\xi &= \varsigma_i \sqrt{D(\Delta v_{\parallel}^2) \Delta t} \\
\Delta v_\zeta &= \varsigma_i \sqrt{\frac{1}{2} D(\Delta v_{\perp}^2) \Delta t} \\
\Delta v_\eta &= \varsigma_i \sqrt{\frac{1}{2} D(\Delta v_{\perp}^2) \Delta t},
\end{aligned}
\tag{3.11}
$$

transforming to the cluster center–of–mass coordinates, we find directly the random fluctuations in velocity,

$$
\begin{aligned}
\Delta v_x &= \Delta v_\xi \frac{v_x}{v} + \Delta v_\eta \frac{(\mathbf{r} \cdot \mathbf{v}) v_x - v^2 x}{v \sqrt{r^2 v^2 - (\mathbf{r} \cdot \mathbf{v})^2}} + \Delta v_\zeta \frac{y v_z - z v_y}{\sqrt{r^2 v^2 - (\mathbf{r} \cdot \mathbf{v})^2}} \\
\Delta v_y &= \Delta v_\xi \frac{v_y}{v} + \Delta v_\eta \frac{(\mathbf{r} \cdot \mathbf{v}) v_y - v^2 y}{v \sqrt{r^2 v^2 - (\mathbf{r} \cdot \mathbf{v})^2}} + \Delta v_\zeta \frac{z v_x - x v_z}{\sqrt{r^2 v^2 - (\mathbf{r} \cdot \mathbf{v})^2}} \\
\Delta v_z &= \Delta v_\xi \frac{v_z}{v} + \Delta v_\eta \frac{(\mathbf{r} \cdot \mathbf{v}) v_z - v^2 z}{v \sqrt{r^2 v^2 - (\mathbf{r} \cdot \mathbf{v})^2}} + \Delta v_\zeta \frac{x v_y - y v_x}{\sqrt{r^2 v^2 - (\mathbf{r} \cdot \mathbf{v})^2}}.
\end{aligned}
\tag{3.12}
$$

The trajectory of the binary was integrated in the cluster center–of–mass frame, using a 4th order Runge–Kutta integrator with quality control. The integrator only integrated the smooth force components, $\nabla \Psi$ and $\mathbf{a_{dyf}}$; the contributions from the random kicks were added after each integration step. With the random kicks added, the quality control on the integrator need not be very stringent, which shortens integration time significantly. To check the accuracy of the integrator, it was run with both dynamical friction and kicks set to zero, and the stability of orbits in the cluster was confirmed; quality control was set to be sufficient to prevent any drift in the orbits. The integration time scale follows naturally from the units selected, $t_{scale} = r_0/\bar{\sigma}$; typical integration times were $10^{10}$ years, requiring $\sim 10^5/t_n$ integration steps or more. The time step used was variable,

$$\Delta t = \epsilon t_n \Big(\frac{1.0 + r}{0.1 + v}\Big), \tag{3.13}$$

where $\epsilon \approx 0.1$, and $t_n \geq 1$ is a time scaling factor, used to allow faster integration by integrating "super–orbits" rather than real orbits. If $t_n \neq 1$ the $\nabla\Psi$ contribution to the force was integrated as if $t_n = 1$, with the contribution due to dynamical friction and kicks scaled as $t_n$ and $\sqrt{t_n}$ respectively. The assumption is that each orbit is representing an average over $t_n$, orbits with perturbations scaled appropriately. In order to integrate a sufficiently large sample of binaries for a sufficiently long time, we chose $t_n \sim 1 - 100$. Care must be taken with $t_n$ large, or the kicks become large compared to the smooth force components. In practice, with a normal distribution of kicks, a few binaries were kicked into escape trajectories during runs with $t_n \gg 10$, so an additional requirement that $\Delta v_{x,y,z} < \max\{v/5, \delta\}$ was added, effectively truncating the normal distribution of $\varsigma$. The truncation could not be proportional to $v$ for very small $v$, lest heavy binaries freeze in the core, after settling by dynamical friction, which is unphysical, and causes numerical pathologies. The truncation of the kick distribution is physically acceptable, as the central limit theorem ensures we will recover the correct normal distribution of kicks by multiple truncated kicks. Binaries could still escape through a succession of kicks, or, by recoil from encounters leading to the binary hardening substantially. A fraction of the binaries on radial orbits were kicked onto trajectories beyond the half–mass radius, and the pericenter of the orbit then kicked out to several core radii before dynamical friction could reduce the apocenter significantly, at which point the relaxation time for the trajectory was typically longer then the integration time (see eg. Sigurdsson 1993).

## *3.2. Encounters*

At each step of the binary's trajectory, the probability of an encounter with a field star, mass $m_\alpha$, $P_\alpha(r,t)$, was evaluated. To calculate $P_\alpha$, we integrated over the local field star distribution distribution, $f_\alpha(r,v)$, calculating the probability that a field star is on a trajectory with pericenter $p$ relative to the binary's center of mass. We say an encounter has occurred if $p \leq sa$ for some value $s\, (= C + D(1+e),\ D = 0.6, C = 4)$ (Hut and Bahcall 1983). An encounter is specified by the pericenter, the relative velocity at infinity between the binary and the field star, and the phase angles of the binary and the field star relative to the binary axis. An encounter was selected by picking a random number, uniformly distributed on $[0,1]$, and comparing it with $P = \sum_\alpha P_\alpha$. If $P$ was greater then or equal to the random number, an encounter was deemed to have occurred. Calculating $P_\alpha$ is the most computing intensive task in the simulation. $v(r)$ is not uniquely defined, due to the varying angular momentum of the trajectory from dynamical friction and kicks, and hence $P(r,v)$ must be calculated at each point in the trajectory. Tabulation and interpolation of $P(r,v)$ was considered, but as the integral could be evaluated in closed form for the distribution chosen, it was more economical to evaluate it explicitly. For more general distribution functions it would probably be better to tabulate $P(r,v)$, and only calculate the partial integrals as needed. To calculate $P$, we calculated the rate of encounters, $R$,

$$\begin{aligned} R(r,v) &= \sum_\alpha \int n_\alpha(r)\sigma(\mathbf{v},\mathbf{v}_*)|\mathbf{v}-\mathbf{v}_*| f_\alpha(\mathbf{v}_*)\, d^3\mathbf{v}_* \\ &= \sum_\alpha n_\alpha(r) I_\alpha, \end{aligned} \tag{3.14}$$

with encounter cross–section, $\sigma(\mathbf{v},\mathbf{v}_*)$, given by

$$\sigma(\mathbf{v},\mathbf{v}_*) = \pi(sa)^2 + \frac{2\pi G(m_b + m_\alpha)(sa)}{|\mathbf{v}-\mathbf{v}_*|^2}, \tag{3.15}$$

and

$$P(r,v) = \Delta t R(r,v) \qquad \text{for } \Delta t R(r,v) \ll 1. \tag{3.16}$$

Requiring $\Delta t R(r,v) \ll 1$ provides an additional time step constraint on the integration of the orbit in the cluster, and can be important for (short lived) wide binaries in dense cluster. Assuming isotropic velocity distribution, we find

$$\begin{aligned} I_\alpha &= \int_0^{v_\alpha} \sigma(\mathbf{v},\mathbf{v}_*)|\mathbf{v}-\mathbf{v}_*| f_\alpha(v_*) v_*^2 \, dv_* \\ &= \frac{8\pi^2}{3}(sa)^2 v^2 \Big[I_1 + I_2\Big] + 8\pi^2 G(m_b + m_\alpha)(sa)\Big[I_3 + I_4\Big] \end{aligned} \tag{3.17}$$

where $I_i$ are given by

$$\begin{aligned} I_1 &= \int_0^{\min\{v,v_\alpha\}} \frac{v_*^2}{v}\left(3 + \left(\frac{v_*}{v}\right)^2\right) f_\alpha(v_*) \, dv_* \\ I_2 &= \Theta(v_\alpha - v) \int_v^{v_\alpha} v_* \left(1 + 3\left(\frac{v_*}{v}\right)^2\right) f_\alpha(v_*) \, dv_* \\ I_3 &= \int_0^{\min\{v,v_\alpha\}} \frac{v_*^2}{v} f_\alpha(v_*) \, dv_* \\ I_4 &= \Theta(v_\alpha - v) \int_v^{v_\alpha} v_* f_\alpha(v_*) \, dv_*, \end{aligned} \tag{3.18}$$

where $\Theta(x)$ is the Heaviside function,

$$\Theta(x) = \begin{cases} 0, & \text{if } x < 0; \\ 1, & \text{if } x \ge 0. \end{cases} \tag{3.19}$$

For the isotropic Michie–King model we can calculate $I_i$ analytically. The integration was carried out using Mathematica, with the results checked using Macsyma and by hand. Defining

$$\begin{aligned} \mu_\alpha &= \frac{m_\alpha}{\bar{m}_c} \\ C_\alpha &= \frac{n_{0\alpha} {\mu_\alpha}^{3/2}}{(2\pi)^{3/2}} \\ W(r) &= \frac{\Psi(r)}{\bar{\sigma}^2} \end{aligned} \tag{3.20}$$

and note that $\mu_\alpha W(r) = \Psi(r)/\sigma_\alpha^2$, then with $v_0 = \min\{v, v_\alpha\}$, we obtain

$$\begin{aligned} I_1 &= \frac{C_\alpha}{v^3}\Bigg[3\sqrt{\frac{\pi}{2}}\frac{1}{{\mu_\alpha}^{5/2}} \mathrm{erf}\left(\sqrt{\frac{\pi}{2}} v_0\right) e^{\mu_\alpha W(r)} \\ &\quad - \frac{v_0}{\mu_\alpha}\left(v_0^2 + \frac{3}{\mu_\alpha}\right) e^{\mu_\alpha W(r) - \frac{1}{2}\mu_\alpha v_0^2} - \frac{v_0^5}{5}\Bigg] + 3I_3 \\ I_2 &= \Theta(v_\alpha - v) C_\alpha \Bigg[\frac{v^2}{4} - \frac{v_\alpha^4}{4v^2} + \frac{1}{\mu_\alpha v}\left(\frac{2}{\mu_\alpha v} + v\right) e^{\mu_\alpha W(r) - \frac{1}{2}\mu_\alpha v^2} \\ &\quad - \frac{1}{\mu_\alpha v}\left(\frac{2}{\mu_\alpha v} + \frac{v_\alpha^2}{v}\right) e^{\mu_\alpha W(r) - \frac{1}{2}\mu_\alpha v_\alpha^2}\Bigg] + I_4 \\ I_3 &= \frac{C_\alpha}{v}\left[\sqrt{\frac{\pi}{2}}\frac{1}{{\mu_\alpha}^{3/2}} \mathrm{erf}\left(\sqrt{\frac{\pi}{2}} v_0\right) e^{\mu_\alpha W(r)} - \frac{v_0}{\mu_\alpha} e^{\mu_\alpha W(r) - \frac{1}{2}\mu_\alpha v_0^2} - \frac{v_0^3}{3}\right] \\ I_4 &= \Theta(v_\alpha - v) C_\alpha \left[\frac{v^2}{2} - \frac{v_\alpha^2}{2} + \frac{1}{\mu_\alpha}\left(e^{\mu_\alpha W(r) - \frac{1}{2}\mu_\alpha v^2} - e^{\mu_\alpha W(r) - \frac{1}{2}\mu_\alpha v_\alpha^2}\right)\right]. \end{aligned} \tag{3.21}$$

At each integration step, $P(v_\alpha = \sqrt{2\Psi(r)})$ was calculated. If an encounter was deemed to have occurred, then $P_\alpha(v_\alpha)$ was evaluated as a function of $v_\alpha$. Which mass group the encounter involves was determined by comparing the fractional probability of $P_\alpha(v_\alpha = \sqrt{2\Psi(r)})$ with the total probability of the encounter taking place, and then $v_\alpha$ was determined by comparing the fractional probability of encounter taking place at different $v_\alpha$, for that mass group.

Given $v_\alpha$, we chose the relative velocity, $v_\infty = |\mathbf{v} - \mathbf{v}_*|$, and hence the impact parameter such that $p \le sa$. We define $\chi$ to be the angle between $\mathbf{v}_\alpha$ and $\mathbf{v}$,

$$\cos\chi = \frac{\mathbf{v}\cdot\mathbf{v}_\alpha}{vv_\alpha}. \tag{3.22}$$

For the isotropic Michie–King distribution function, an analytic expression for $\chi \in [0,\pi]$ can be found. We picked a random number $\varsigma_i$, uniform on $[0,1]$, and chose $\chi$ from

$$\varsigma_i = \frac{R(\chi|m_\alpha, v_\alpha)}{R(\pi|m_\alpha, v_\alpha)}. \tag{3.23}$$

Integrating, and defining some auxiliary variables,

$$\begin{aligned} c_0 &= |v - v_\alpha| \\ c_1 &= |v + v_\alpha| \\ \beta &= \frac{2G(m_b + m_\alpha)}{sa} \\ \gamma &= -c_0(1-\varsigma_i)(c_0^2+\beta) - c_1\varsigma_i(c_1^2+\beta), \end{aligned} \tag{3.24}$$

we find after some algebra, that $\chi$ satisfies

$$t^3 + \beta t + \gamma = 0, \tag{3.25}$$

where $t = \sqrt{v^2 + v_\alpha^2 - 2vv_\alpha\cos\chi}$, and hence

$$\cos\chi = \left[\frac{(A+B)^2 - (v^2+v_\alpha^2)}{-2vv_\alpha}\right], \tag{3.26}$$

where $A+B$ is the solution of the cubic,

$$\begin{aligned} t &= A + B \\ A &= \sqrt[3]{-\frac{\gamma}{2} + \sqrt{\frac{\gamma^2}{4} + \frac{\beta^3}{27}}} \\ B &= -\sqrt[3]{\frac{\gamma}{2} + \sqrt{\frac{\gamma^2}{4} + \frac{\beta^3}{27}}}. \end{aligned} \tag{3.27}$$

Knowing $v_\infty$, and the maximum pericenter, we calculated the maximum impact parameter, adjusted for gravitational focusing. Given the maximum impact parameter, we picked the actual impact parameter, distributed uniformly in the area of the beam provided by the maximum permitted impact parameter. The beam is symmetric about the axis between the binary and the field star, and a phase angle for the approaching field star was picked at random, as was the angle of the binary axis relative to the axis joining the stars. Given the field star mass group, the field star stellar type and radius $R_{*_3}$ were set, using the same algorithm used to select the radii and stellar type of the binary members. It should be noted that the collision probability per integration step was typically less then the range of built–in random number generators in most computers ($1/(2^{31}-1)$ for DEC3100s), and to get a reliable encounter rate the random number range must be extended. We used a uniform conditional probability distribution providing a smooth distribution to less then $10^{-14}$, which was sufficient for our purposes. If the random number chosen was less then $10^{-7}$, the probability was scaled up by $10^7$, and a new random number was drawn uniformly on the interval $(0,1)$, and compared with the scaled encounter probability. This provides independent uniform sampling to less then $10^{-14}$, provided the random number generator has no sampling correlations.

Having chosen an impact parameter, relative velocity and phase angle, the encounter was integrated explicitly, using the three–body integration scheme described in Sigurdsson & Phinney (1993). The binary parameters and the relative velocity were scaled to units where $a = 1$, and the three–body trajectory was calculated explicitly until resolved, or the number of integration steps exceeded a fixed maximum ($= 2 \times 10^6$ steps; previous calculations indicated that a very small proportion of encounters required more then $10^6$ steps). Every 20,000 integration steps the state of the system was checked to see if the encounter was resolved. Explicit provision was made for tidal encounters leading to a merger, merger being assumed if the separation between any pair of stars was less then $f_t \times (R_{*_i} + R_{*_j})$, $f_t \approx 3.1$ (Lee and Ostriker 1986, see also Benz and Hills 1992), $f_c$ was chosen less than $f_t$; this was deliberate in anticipation that an encounter with a contact binary ($a \leq f_t(R_{*_1} + R_{*_2})$) would perturb the binary sufficiently for the system to undergo an energetic event. If a merger occurred, the orbital parameters of the resulting system were calculated, assuming an impulsive merger of the two stars, and no mass loss. The stellar type and radius of the merged star were also determined. If the resulting system was bound, the new binary was returned to the cluster for further integration. After each encounter was complete, the final state binary was returned to the main integrator for further integration. If no binary existed after the encounter (system was ionized, or merged leaving the third star unbound) the run was halted and a new binary was picked. If a binary was available for integration, its position in the cluster was updated assuming linear extrapolation of the binary's pre–encounter velocity in the cluster center–of–mass frame, and the velocity in the cluster was adjusted to allow for the outcome of the interaction. For very soft binaries on orbits well outside the core, the encounter rate was dominated by softening encounters with the lowest mass stars. To avoid spending excessive computing time on these gradual ionizations, a binary was arbitrarily considered to be effectively ionized if its semi–major axis exceeded $1.2 \times \max\{a_i, a_{max}\}$, with $a_{max} = 10\,AU$ for most models, as ionization for binaries that wide is virtually inevitable, and collisions during encounters are very unlikely for a binary that wide. The integration of the binary trajectory in the cluster center–of–mass was then continued until the orbit integration had been run for time $T$, at which point the final state of the binary was saved and a new binary was picked. A typical run consisted of $100 - 1000$ binaries, with $T = 5 \times 10^8 - 10^{10}$ years, less concentrated clusters being evolved for longer times.

# 4. Results

We have run simulations using over 20 different models of globular clusters. Here we discuss some of the general results derived from 13 of those models, involving some 15,000 binaries in 34 runs, where a "run" describes the time evolution and interaction of a set of binaries in a particular model for time $T$. Other simulations and results are described elsewhere and focus on modeling particular phenomena in (particular) globular clusters (eg. Sigurdsson 1993, Sigurdsson, Davies & Bolte 1994). The rates derived here are for binary–single star processes only, (T3 in the notation of Phinney & Kulkarni [1994]), the relative contribution of other processes will be considered in the next section.

The expected number of products of various processes, $X$, $\Gamma(X)$, is estimated from,

$$\Gamma = f_b(f_r)f_w \frac{n_R}{N_b} N_* \frac{\tau}{T}, \tag{4.1}$$

$n_R$ is the number of candidate products observed out of $N_b$ simulations computed, $M_T$ is the total mass of stars in the cluster as shown in Table 2, $f_r$ is the retention rate for the stellar types considered, assumed unity for all except neutron stars, and $\tau$ is the mean lifetime of the observable result of the process (eg, pulsar or blue straggler), compared to the duration of the simulation, $T$.

$f_b$ is the binary fraction of the cluster. Our default assumption is that primordial binaries are distributed uniformly in $\log(a)$, with 10% in each decade in $a$. That is, 50% of the stars are binaries with semi–major axis between $10^{-2}$ and $10^3\,AU$. Wider binaries are ionized shortly after the formation of the cluster, tighter binaries are liable to spiral in to contact. $f_w$ is the binary weight, as shown in Table 3, it represents the fraction of binaries in the mass–range chosen for the run, assuming the secondary IMF is drawn independently from the global IMF. $\Gamma$ is readily estimated for other choices for $f_b$, $f_r$, $\tau$. Correcting for $f_w$ is non–trivial, as the encounter cross-section is strongly dependent on the masses of the stars interacting. $f_w$ ideally should be corrected for the changes in the binary mass function due to exchanges, especially for the denser clusters. $f_b$ may also be corrected for the change in the primordial binary distribution as the cluster evolves. In particular, hardening of binaries due to encounters early in the evolution of the cluster may lead to an excess of binaries with semi–major axis, $a_c$, such that $\sigma(a_c)n_0\bar{\sigma}(0) \sim t_r^{-1}$, where $t_r$ is the cluster relaxation time scale and $\sigma(a_c)$ is the cross–section for hardening encounters for a binary with semi–major axis $a_c$.

## 4.1. *Cluster Properties*

The thirteen cluster models discussed here are divided into five classes. Two classes of low density clusters, two classes of high density clusters and a broad class of intermediate density clusters of varying concentration. To compare the cluster models with real clusters, it is best to compare the core radius, $r_c$, the surface luminosity density $\Sigma_L(0)$ and the ratio of the half–mass radius, $r_h$, to the core radius. The model luminosity density follows from the relation given by Djorgovski (1993),

$$\Sigma_L(0) = \frac{p\bar{m}_c n_0 r_c}{(M/L)}, \tag{4.2}$$

where $p \approx 2$ is the concentration correction given by Djorgovski and $M/L \sim 2$ is the mass–to–light ratio. Ideally $M/L$ is uniquely determined by the IMF, but we leave it as a free parameter as other model parameters have large systematic uncertainties that are degenerate to variations in $M/L$.

The models considered here (Table 2) are mostly motivated by real clusters containing recycled pulsars (Taylor, Manchester & Lyne 1993), with the exception of the low density model 1, which was considered as broadly representative of the many low density, low dispersion globular clusters in the halo. As a first approximation, models 2 can be considered representative of clusters like M53 and M13; models 3 are comparable to M4 and M5; models 4 are an attempt to represent 47 Tuc or Ter 5, and model 5 is a model of M15. Core–collapsed clusters are not well represented by any Michie–King model, and our model 5 attempts to fit the core and inner parts of the cluster only.

The choice of model parameters was partly driven by theoretical biases. The choice of IMF follows Chernoff & Weinberg (1990) as does the choice of distribution of evolved remnants from higher mass stars. The relatively flat Salpeter IMF is partly justified by observational constraints (Phinney 1992, Pryor [private communication]), although its extension to both the low and high end of the mass range is poorly constrained observationally. At the low end, observations are confounded by mass–segregation biasing the locally observed IMF (see eg. Pryor *et al.* 1989) and differential mass–loss due to tidal stripping; at the high end, there is no direct data, but comparisons can be made with the mass function of open clusters and theoretical models which are consistent with the IMF chosen here (Murray [private communication]). Ultimately, it is the number of dark, massive degenerate remnants that is important. In the model this is determined both by the IMF and the evolutionary model for the remnant mass as a function of progenitor mass. Our models have a large dark remnant fraction, both neutron stars (NSs) and massive white dwarfs (HWDs), other choices are possible (cf. Meylan 1989). Our choice is partly dictated by the need for a large number of NSs or HWDs to produce the observed MSPs and LMXBs, partly by observations of high $M/L$ in concentrated clusters (Phinney 1992, Pryor [private communication]). Dynamically all that matters is the current number fraction of dark remnants in the core, independent of their progenitor mass distribution.

The NS fraction is further confounded by the fact that it seems likely that most NS are born with "kicks" and recoil at high speeds relative to their local standard of rest (Lyne & Lorimer 1994), and such NS could not remain bound to a globular cluster. Clearly some NS are retained, with estimates as high as 30% (Hut & Verbunt 1983). The retention factor is determined by the low end tail of the kick distribution which is poorly determined observationally. Further, it is not clear that the kick distribution of the currently forming NS is comparable to that of the low metallicity progenitors of the globular cluster NS. We draw some comfort from the observation of PSR J1713+0747 (Camilo, Foster & Wolszczan 1994) which is a low speed binary pulsar in the galaxy and would most likely have remained bound to a globular cluster if born in one. We correct for any discrepancy between the model number of NS and those actually retained in cluster with a *post hoc* correction factor, $f_r$, when estimating formation rates.

It is worth noting some features of the models chosen. With multi–mass models, the concentration is not a unique function of the potential depth, $W_0$. The core radius is determined from the surface density profile of the turnoff mass stars. The turnoff mass, $m_t$, need not be the same as the mean core mass which determines the dynamical scale radius, $r_0$. If $m_t > \bar{m}_c$ then $r_c < r_0$ and vica versa. Steeper mass functions lead to smaller numbers of NSs and HWDs and thus a lower $\bar{m}_c$, while low $W_0$ clusters are less mass–segregated and have a lower $\bar{m}_c$. The net effect (see Figure 1) is that multi–mass models span a smaller concentration range as a function of $W_0$. In particular we are forced to higher $W_0$ to model concentrated clusters. It should also be noted that the dispersion, $\bar{\sigma}(0)$, used in the models is the intrinsic 1–D dispersion of an (imaginary) ensemble of stars of mass $\bar{m}_c$, the projected 1–D dispersion of the turnoff mass stars is the observable quantity and is typically $\sim 20 - 30\,\%$ smaller except for the most concentrated models.

As discussed in Section 2, the code has been adapted to generate N–body realizations of multi–mass Michie King models. We are using these as initial conditions for N–body modeling of mass–loss in globular clusters and tidal shocking (Sigurdsson & Hernquist in preparation). These N–body realizations are also potentially useful for Monte–Carlo modeling of observations of actual globular clusters. The code may be obtained from the first author.

## *4.2. Encounter Probabilities*

The encounter probability for binaries (and single stars) on different orbits in different clusters are of intrinsic interest. In addition to the total encounter probability, we can quantify the relative probability of encountering different mass stars, and the velocity distribution of encounters (see also Heggie & Hut 1993). The total encounter probabilities calculated for different models were compared with analytic estimates and found to be in agreement.

Of particular interest is the relative probability for a binary to encounter different mass stars as a function of cluster parameters and mass function. The encounter probabilities are necessarily dominated by encounters in the core, and we consider the fractional, orbit averaged relative encounter probability for a binary moving across the core. That is, the partial integral of the probability of encountering a star from mass–group $\alpha$ as a function of radius as the binary moves through a (half)orbit from pericenter to apocenter, normalized to the total encounter probability over the (half)orbit, $P_{tot}$. Some representative results are shown in Figures 2a–2d.

The relative encounter probability depends both on the mass and dispersion of each mass–group, and the core density of the stars of that mass. The gravitational focusing strongly biases encounters towards high mass, low dispersion stars, while a steep mass function and low concentration provides a high core density of low mass stars. For all models with $x_* \geq 1$ the encounter probability outside the half–mass radius is dominated by the lowest mass stars, simply by virtue of their total number and the depletion of high mass stars from the halo by mass–segregation.

The relative probability of encounters with stars near the turnoff (mass groups 7 and 8 for models with 10 mass groups), compared to the probability of encountering massive dark remnants (mass groups 9 and 10), is critical to predicted observational properties of exotic stellar objects in globular clusters. In the low concentration clusters, encounters with turnoff mass stars are a significant fraction of the total encounter rate in the core (see Figure 2a), with a comparable contribution from the lowest mass stars, especially near the edge of the core. For moderate concentration models, such as model 3.4 (see Figure 2b), the turnoff mass stars dominate the encounter rate in the core, with an encounter with a turnoff mass star being twice as likely as with a HWD and three times more likely than with a NS. For $W_0 \sim 6-9$ the relative encounter probabilities are sensitive to the mass function, for a model with a slightly flatter mass function the encounter rate would be dominate by the NS (mass group 10), while a steeper mass function than Salpeter would ensure the encounter rate would be dominated by turnoff mass stars over a broad range of intermediate concentration clusters.

For the most concentrated models (see Figures 2c, 2d) the encounter rate in the core is completely dominated by the NSs and HWDs, unless the mass function is very steep. For our model 4.2, which approximates a cluster like 47Tuc, over 70% of binary–single star encounters in the core should be with a massive degenerate, if our model is at all representative of the true mass function in the cluster. As we discuss later, this implies the relative frequency of blue stragglers, CVs and MSPs should vary predictably with cluster parameter. For model 5, our model for M15, the core encounters are completely dominated by the neutron star population (mass group 10). This is in accordance with observations, which suggest a significant population of NSs and HWDs in the core of M15 (Phinney 1992) and the observation of 10 MSPs in M15 (Anderson *et al.* 1990, Anderson 1992). The fractional encounter rate for turnoff mass stars in the core of M15 is only 2.5% according to the model. Outside the core, the turnoff mass stars dominate the encounter rate for binaries on orbits at $5-10\,r_c$ with the lowest mass stars dominating by the half–mass radius. As the density at few $r_c$ is $\sim 10^4$ pc$^{-3}$ the number of encounters outside the core can be significant in M15.

The domination of the encounter rate by low mass stars outside cluster cores is important, as the critical velocity, $v_c$, that determines the "hardness" of a binary is $\propto 1/\sqrt{m_3}$, where $m_3$ is the mass of the single star encountering the binary. *Binaries that would be ionized in the core are hardened in the halo.* This is in contrast with single–mass models in which ionization of wide binaries in the halo is efficient, whereas in the multi–mass models the wide binaries may harden through encounters with low mass stars in the outskirts of the cluster, and become "hard" by the time they reach the core through dynamical friction, where they have an appreciable chance of encountering the more massive stars. Thus the "hardness" of a binary is a function of both the local dispersion and the nature of the local background population of field stars.

Encounter probabilities for binaries or stars on different orbits are readily calculated for any multi–mass Michie–King distribution. Calculations for particular models or families of models are available by arrangement with the authors.

## 4.3. *High Energy Stars*

The orbital evolution portion of the code allows modeling of mass loss from globular clusters. Our results agree with those of Johnstone (1993), with a few percent of the stars drifting across the tidal radius per half–mass relaxation time, and the mass loss biased to low mass stars. As Johnstone found, the dominant mechanism appears to be orbit perturbations near periastron for stars with high (orbital) energy on highly radial orbits. As noted by Lee & Ostriker (1987), a star is not necessarily lost to the cluster if it crosses the tidal radius, its orbit will still penetrate the cluster and further perturbations due to the inhomogeneity of the cluster potential may drop the apastron inside the tidal radius before galactic tides unbind the star. This is more likely for clusters on radial orbits about the galaxy, which spend most of their time at large galactic radii, the tidal radius being effectively determined by the galactic tides at perigalacton.

High energy stars are present in globular clusters, both as a consequence of the natural tail of the truncated Maxwellian velocity distribution, and because of recoil from stellar encounters in the core. Encounters involving binaries are particularly effective in generating a population of high energy stars (Hut & Bahcall 1983, Sigurdsson & Phinney 1993). If a star is ejected from the core of globular cluster, and remains bound to the cluster, its orbit evolves under the combined effects of dynamical friction and the "Brownian kicks". Dynamical friction acts to lower the energy of the star, while the "kicks" can nudge the periastron of the orbit out of the core, leading either to escape from the cluster, or a moderately eccentric halo orbit. As the star spends most of its time near apastron, where relaxation time scales are long, the relaxation of the orbit is quite slow, even for concentrated clusters, and we may ask whether such high energy stars could be observed. This is particularly interesting in view of the high velocity interlopers observed in 47Tuc (Meylan, Dubath & Mayor 1991) and other clusters (Pryor, Peterson [private communication]).

A star can only be observed with a large projected velocity near the core of a cluster. High energy bound stars at large radii are necessarily moving at low speeds, and only when passing through the core can they be seen, in projection, moving at high speeds. Two factors confound the interpretation of observations of high speed stars; there is a tendency to reject them as cluster members because their radial velocity is far from cluster mean, and there is an ambiguity as to whether the core dispersion is to be calculated from a sample including the high velocity stars or not. Clearly including high velocity stars in the sample from which the dispersion is derived, leads to a higher derived dispersion. Theoretically, one expects any cluster member observed to be bound to the cluster, to see a star in the core in the process of being ejected is *a priori* very unlikely unless there is rapid mass loss from the cluster through stellar ejection. Thus the high velocity stars in 47Tuc should be considered to constrain the central escape velocity of the cluster, if they are in fact in the cluster they must be bound to the cluster.

An interesting question to ask is how probable is it that we might see high energy stars ejected from the cluster core, at high projected velocities near the core, and how that compares to the high velocity tail of the intrinsic velocity distribution? We did Monte Carlo simulations of the projected velocity of stars ejected from the core of clusters at 90% of the core escape velocity, tracking the projected velocity, $v_p$, as a fraction of the (intrinsic) core dispersion, $\bar{\sigma}$. The probability of observing the star at high projected velocity is time dependent, as relaxation takes effect and the orbit circularizes in the cluster, it becomes less probable we see the star at high projected velocity. Figure 3 shows the fraction of time a 1 $M_\odot$ star ejected at $0.9\sqrt{2W_0}\bar{\sigma}$ will be observed at different projected velocities, averaged for $10^8$ years and $10^9$ years of orbit evolution, for three clusters with different $W_0$. For a 47Tuc class cluster ($W_0 = 12$), orbit evolution is not very strong even over $10^9$ years as the star spends a large fraction of its time in the cluster halo. The probability of observing a star at $3 - 4\bar{\sigma}$ is approximately $5 \times 10^{-3}$, requiring a total population of few hundred such stars in order for two to be observed at this time. If the high velocity stars in 47Tuc are representative of a set of stars ejected from the cluster core through binary interactions, we infer that $\sim 500$ such interactions must have taken place in the last $10^9$ years. Both the stars observed are above the main sequence turnoff, however many of the binary collision scenarios for producing such stars could lead to the ejected star moving off the main sequence (Davies *et al.* 1991, 1994, Sigurdsson *et al.* 1994).

### *4.3.1. Binary Neutron Star Ejection*

In very dense clusters, NS–NS binaries form and may be hardened at high enough a rate to have a significant probability of being ejected from the cluster core, or even beyond the cluster tidal radius. Such a system is observed in M15, PSR B2127+11C (Anderson *et al.* 1991, Anderson 1992, Phinney & Sigurdsson 1991). Hard NS–NS binaries are of interest as sources of gravitational radiation for planned observatories like LIGO, and as sensitive tests of General Relativity when observed directly (Taylor & Weisberg 1989, Taylor 1992). Of the three observed NS–NS binaries, one is in a globular cluster. A factor in estimating the total rate of NS–NS binary formation (and subsequent merger rate) is what fraction remain bound to the cluster compared to the number ejected out of the cluster and counted as part of the field population (Phinney 1991). We find that NS–NS binary ejection is only important in the very densest clusters, such as M15, which are thought to have undergone core–collapse. While NS–NS binaries are formed in lower density clusters such as 47Tuc, they are not hardened to the point of ejection. There were no NS–NS binaries in our original binary distribution, all formed by exchange or collision.

We find that $\sim 20\%$ of the NS–NS binaries formed in our densest cluster are ejected from the cluster, with another 15% ejected to the cluster halo. Half the ejected NS–NS binaries were formed in "clean" exchanges where there would not have been any opportunity for mass–transfer onto either neutron star.

Such systems would presumably not be detectable as radio sources. Of the NS–NS binaries remaining in the core (or ejected to small radii where dynamical friction rapidly returned them to the core), 15% were involved in a collision that led to a "soft" final binary and one would expect the resulting pulsar to be observed as a single pulsar, its neutron star companion being rapidly unbound from the system by subsequent encounters. None of the remaining 50% NS–NS binaries were tagged as having had an opportunity for mass–transfer, all were in hard binaries, but not hard enough to undergo spiral–in and merger though gravitational radiation. It would be expected that these binaries would undergo further encounters on time scales of $10^9$ y with further ejections taking place as the surviving NS–NS binary population hardened or collided. Interestingly, no NS–PSR binary was inferred to be observable in the cluster cores. Statistically one infers there should be 2–3 NS–PSR binaries in the cluster halo for any in the core.

## 4.4. *Heating*

Binary interactions in the core of a globular cluster can heat the core and stave off core–collapse for dense clusters (Goodmand & Hut 1989). We estimated heating rates due to the binary interactions. A presentation of some of the detailed model results is published in Sigurdsson (1991) [see also Sigurdsson & Phinney 1993]. The results are sensitive to the assumptions made about the initial binary semi–major axis distribution, and only qualitative conclusions can be made about general results. Net heating is also affected by the mass loss due to ejections from the cluster core and energy dissipation during stellar collisions.

In the low density clusters, binary encounters produce net cooling, of order $10^{-2}\,kT$ per binary. This is simply because our initial binary distribution included a fraction of soft binaries that were rapidly ionized during their orbital evolution in the cluster. For the higher density clusters, heating of $\sim 1-10\,kT$ per binary was observed with the exact value subject to systematic variation by an order of magnitude depending on the initial conditions chosen. Clearly, hard, massive binaries concentrated in the core of a globular cluster can be efficient sources of heating. Reliable quantification of the heating rate can be made with our method, but should be done on an individual cluster basis.

## 4.5. *Binary Distribution*

There is strong evolution of the radial distribution of binaries within a globular cluster, due to dynamical friction, the Brownian kicks, and encounters.

We tested the code by generating populations of binaries with a non–equilibrium distribution, and allowed them to evolve without encounters taking place. Binaries that were initially distributed with a under–relaxed distribution relaxed to the correct equilibrium distribution for point masses of the same distribution, on a relaxation time scale. We also over–relaxed an ensemble of binaries, that is we evolved a population of low mass binaries with an initial concentration appropriate to a higher mass population; the final distribution relaxed out to the correct, less concentrated equilibrium distribution appropriate to a population of the mass chosen.

Some representative binary radial distribution profiles are shown in Figure 4a–f. The Figures show typical distributions of binaries for globular clusters of different concentration and density. The plots show the number of binaries per octave in radius for the initial and final populations. The last bin shows binaries with excursions beyond the tidal radius either through Brownian kicks or ejections due to encounters.

Figure 4a shows the relaxation of an over–relaxed binary population. The binary population included a large fraction of low mass binaries that were over concentrated for their mass. The final distribution shows the relaxed binary distribution after $10^{10}$ years. Encounters were not significant in modifying the distribution, about 7% of the binaries were lost to the cluster, predominantly the lowest mass binaries. Figure 4b shows the evolution of a population of more massive binaries in the same cluster. The binaries now relax to a more concentrated distribution appropriate to their mass. Significantly fewer binaries are lost from the cluster.

Figure 4c shows the evolution of different binary sub–populations in a "typical" moderately concentrated, moderately dense ($n_0 = 10^4$) globular cluster. The top two panels are for a more massive binary population, while the two left panels are for a initially less concentrated binary population. The less concentrated populations spend most of the time relaxing to the core and the net encounter rate for those is a factor of 3–4 smaller than for the initially concentrated populations. For the initially concentrated populations,

Models 3.2.4 and 3.2.6, shown on the right in Figure 4c, recoil due to encounters is a significant factor in the evolution of the population. Over 10% of the binaries, mostly those with binding energies in the $1 - 10\,kT$ range, undergo flybys or exchanges during which the binaries harden and recoil from the core of the cluster. As the relaxation time at $r_h$ is $\gtrsim 10^9$ y for these clusters, binaries ejected do not relax rapidly back the core.

Figures 4d and 4e show the relaxation and ejection process for a denser, more concentrated cluster, one comparable to 47Tuc or Ter 5. Again the massive binaries relax to the core and undergo encounters. All the binaries with binding energies in the $1 - 10\,kT$ range that reached the core underwent encounters, as did 20% of the lower mass and 50% of the high mass binaries in the $10 - 100\,kT$ range. There is now a significant overpopulation of binaries ejected to $\sim r_h$, where the binaries go into "parking orbits", as Brownian kicks circularize the orbits outside the core before dynamical friction can bring the binaries back to the core. For Model 4.3, a few percent of the binaries are ejected from the cluster by recoil from encounters.

Figure 4f shows the evolution of sample binaries in our model core–collapsed cluster. The process of relaxation and ejection is even more pronounced than for models 4.1 and 4.3 despite $T$ being order of magnitude shorter. The depletion of binaries in the $1 - 5\,r_c$ region is noticeable in these models. This is the region where the dynamical friction time scale is $\ll T$ and the binaries segregate rapidly to the core. Binaries around $r_h$ do not relax completely on this time scale. The lower left panel shows the evolution of a sample "halo" population of binaries, a population of lower mass binaries considered to have remained at large radii in the cluster as the core collapsed, either because they were formed at large radii, or because they were ejected to the halo at early stages of the cluster evolution as seen in Figure 4c. A few of these wide, relatively low mass binaries diffuse to the core during the simulation, where they rapidly undergo encounters, are hardened and more massive field stars exchange into the binary until it is ejected back to $r_h$ or beyond by recoil from the encounters.

## *4.6. Exchanges*

A key feature of multi–mass binary–single star encounters in globular clusters is the large cross–section for exchange (Heggie 1975, Hills 1976, Hut & Bahcall 1983, Hills & Fullerton 1980, Fullerton & Hills 1982, Hills 1992, Sigurdsson & Phinney 1993). The primary effect of exchanges that we are concerned with is the change in stellar membership of a binary during exchanges, with high mass stars and stellar remnants preferentially exchanged into binaries in evolved, dynamically cool systems, such as the current galactic globular clusters.

It is possible for a small fraction of neutron stars to become bound to a primordial low mass secondary if there is a modest kick on the neutron star during the supernova. Such systems undoubtedly account for many of the galactic low mass binary pulsars with the secondary in wide, near circular orbits. There are several such systems in the galaxy, including PSR B1800-27, PSR B1953+29 and PSRJ 2019+2425 (Taylor *et al.* 1993), consisting of a recycled pulsar and a low mass white dwarf. The white dwarf is, presumably, the remnant of a low mass main–sequence star (MS) that evolved off the main–sequence within the last $\sim 10^9$ years. It is possible that such a system could remain bound to a globular cluster on formation, with PSR J1713+0747 being an example in the galaxy (Phinney & Kulkarni 1994, Bailes 1989).

Two low mass binary pulsars with white dwarf secondaries in wide, low eccentricity orbits are observed in globular clusters, PSR B1620-26 in M4 and PSR B1310+18 in M53. M53 and M4 have densities of $\sim 10^3$ and $10^4\,\mathrm{pc}^{-3}$ respectively. Correcting for observational selection effects, the inferred birthrate per unit mass for low mass binary pulsars in globular clusters is excessively high (Phinney & Kulkarni 1994 and references therein). Exchanges provide a natural way of producing the progenitors of these PSR-WD binaries. Either a single NS may encounter a binary containing a MS near the turnoff and replaced the binary secondary through exchange; or, a NS in a binary with a low mass companion encounters a single or binary MS at the turnoff, with the encounter producing a NS–MS binary. The MS subsequently evolves off the main–sequence, spinning up the NS, leaving a wide orbit, circularized PSR–WD binary. A NS in a low mass binary may either form by exchanging a single NS into a binary with a companion of mass less than the turnoff mass, or may it may be a primordial NS–WD binary, with the WD descended from a moderately massive star, mass $\gtrsim 1.5\,\mathrm{M}_\odot$. Such a binary would have undergone unstable mass transfer when the secondary evolved off the main–sequence, leaving a medium or low mass white dwarf in a tight orbit about the NS. The NS would most likely have been spun–up to a pulsar at the time, but would have spun down past the deathline by now, the original mass transfer necessarily having taken place $\sim 10^{10}$ years ago. Rappaport *et al.* (1990) discuss such a scenario for forming PSR B1620-26 (see also Sigurdsson 1993).

In order for such an exchange to produce a low mass binary pulsar (LMBP), the final NS–MS binary must have a semi–major axis $a \lesssim 0.5\,AU$. Due to the increase in $a$ during exchange, this process is dominated by primordial binaries with $a \sim 0.1 - 0.3\,AU$. In clusters with $n_0 \lesssim 2 \times 10^3\,\mathrm{pc}^{-3}$, such encounters are rare and poorly explored by our method due to the very small number of such encounters in the total population of low density globular clusters. The formation rate in the low density globulars can readily inferred from exchange cross–sections (Sigurdsson & Phinney) given some assumed binary population.

For intermediate density clusters, our simulations should be representative of net rate for such exchange binaries. Model 3.2.4 produced two such exchange binaries, one ended up in the core, the other suffered modest recoil and at the end of the simulation was in an orbit at about $3r_c$. In model 4.1 another two such binaries formed, one of which was ejected from the cluster, the other ejected to few $r_c$. Model 4.2.1 produced 7 such exchange binaries, three of which were ejected to several $r_c$, the remainder ending up in the core. For higher density clusters, such wide binaries have time scales for encounters in the core that are short compared to the observed LMBP characteristic ages.

Model 3.2 is comparable to M4. With 1000 binaries run, then correcting for the mass–range used, these would be representative of a global binary fraction of $\sim 20\%$ in that semi–major axis range, or a total primordial binary fraction of order 50%. The number of exchange LMBPs expected per globular cluster in this class, $\Gamma_{LMBP-ex} = 0.3$ from model 3.2.4. For a beaming factor of 0.5 and pulsar lifetime of $10^9$ years, we expect to see only one LMBP like PSR 1620-26 per six to seven clusters like M4. From this we infer a birthrate for PSR 1620-26 class LMBPs in moderate density clusters of $1 - 2 \times 10^{-10}$ per year. This compares with a total inferred observable PSR birthrate of $5 \times 10^{-10}$ per year in such clusters, for all formation channels (Phinney & Kulkarni 1994).

The situation is more interesting in the denser clusters of $n_0 \sim 10^5$. There the production rate is an order of magnitude higher, but with approximately half the pulsars expected to be seen outside the core near $r_h$. NS–MS binaries that return quickly to the core of those dense clusters should undergo more encounters in the core, particularly after spin–up when the binary has expanded through the conservative mass–transfer. With a final secondary mass of $\lesssim 0.3\,\mathrm{M}_\odot$, an LMBP with $a \sim 0.5\,AU$ is only marginally hard. The LMBP will either ionize on a time scale $\gtrsim 10^9$ y, leaving a single pulsar, or will harden (and probably exchange secondary )on a time scale of $\sim 5 \times 10^8$ years. We predict that a small population of eccentric, $a \sim 0.3\,AU$, LMBPs should be observed in moderately dense clusters ($n_0 \sim 10^5\,\mathrm{pc}^{-3}$), with about equal numbers in the core and halo, the core population having significant eccentricities ($\sim 0.3 - 0.7$), the halo population having small eccentricities, comparable to PSR 1620-26.

## 4.7. *Collisions*

The primary goal of this work was to attempt to quantify binary–single star interactions which lead to the formation of MSPs. As discussed earlier, we require that mass–transfer take place onto a NS (or HWD if AIC occurs) in order for a MSP to be formed. During our integration of the individual encounters, we kept track of the separation between individual stars, and compared the stellar separation to the assigned stellar radii. If an encounter led to an approach within $3.1R_*$ between any pair of stars, we assumed a tidal encounter had taken place and the encounter was flagged as a candidate for mass transfer taking place. In order to investigate the dynamical evolution of such encounters, we assumed the stars involved merged in a completely inelastic collision, conserving momentum only, (see Davies *et al.* 1994) and estimated the relative orbital parameters of the third star involved in the encounter. If the third star was found to be bound to the merged stars in this approximation, the new binary was returned to the cluster for further orbit integration in the cluster potential. If the third star was judged to be unbound from the merged star, orbit integration was terminated. The evolution of the orbits of single merged stars in globular clusters is considered separately (Sigurdsson *et al.* 1994, Davies & Sigurdsson [in preparation]).

Table 4 shows the statistics for collisions in the simulations discussed here. We separate the collisions as to the most massive star involved, $n_{NS}$ counts collisions involving neutron stars, $n_{WD}$ counts collisions involving white dwarves, and $n_{MS}$ counts collisions involving main–sequence and (sub)giant stars only. Assuming that all such collisions lead to the productions of MSPs, CVs and BSs we discuss some of the implications for total and relative production rates for these objects in different globular clusters. One fact readily discerned from Table 4, is that only binaries already concentrated in the core of the cluster are relevant for calculation of collision rates. We therefore assume that (at least for $n_0 \gtrsim 10^4$) the correct rates are to be inferred from the runs for which the binaries are initially relaxed to the core.

A severe problem in deriving statistical production rates for MSPs and LMXBs is not only that the total number derived from our simulations is small, but also that the total number in the different class globular clusters is also small and subject to the actual vagaries of "unique" events. In calculating $\Gamma_{PSR,CV,BS}$ for the respective objects, we will assume an uncorrected $f_b$ of 0.1 per decade in $a$, a $f_r = 0.2$ and $\tau_{PSR} = 10^9$, $\tau_{CV} = 3 \times 10^9$ and $\tau_{BS} = 5 \times 10^9$. We note that MSP lifetimes are variable, and the variation may be systematic with globular cluster parameters as NS may be spun up to different spin periods according to the creation process and (for high density clusters) the local stellar density. The observable lifetime of CVs is sensitive to whether mass transfer is stable or not, and the flux to which observers are sensitive, and $\tau_{CV}$ should be considered for scaling purposes only.

### 4.7.1. *Neutron Stars*

In the low and moderate density clusters we only found one collision involving a neutron star, out of a total of 16 collisions. In Model 3.2.4, a field WD encountered a wide ($a \sim 5\,AU$) NS–MS binary (typical of exchange binaries formed at low densities) and during the interaction the NS and MS collided. Such a collision would presumably lead to a single MSP, possibly with a very low mass remnant core of the MS in a very tight orbit. We infer $\Gamma_{MSP-coll} \lesssim 0.3$. Ignoring the statistical uncertainty, we predict that such processes occur at about half the rate for exchanges leading to LMBPs discussed above.

The situation is more interesting for the higher density clusters, where in addition to a statistically significant $\Gamma$, we see a number of individually interesting collisions in the simulations.

Models 4.1.2,3 provide 4 NS collisions, one from what was initially a NS–MS contact binary. Inferred $\Gamma_{MSP-coll} = 8$. Assuming a beaming factor of 0.5, we infer 4 observable MSPs per cluster. This compares with Model 4.2.1 where we have 8 NS collisions, for a $\Gamma_{MSP-coll} = 2$. One of the collisions was in a binary where mass transfer would likely be taking place prior to the collision. Further, in Model 4.2.1, we see in addition 2 WD–MS collisions where an NS is inferred to be bound to the system in a tight orbit ($a < 1\,AU$); 3 MS–MS collisions in which a companion NS is left bound, one in an orbit where it is likely to collide with the merged stars; and 2 WD–WD collisions which might lead to AIC and hence a MSP. Thus we infer $\Gamma_{MSP} \approx 4$, and $\sim 2$ observable MSPs, given our assumptions. As model 4.2 is only half as massive as Model 4.1, both simulations lead to an estimated 2 observable MSPs per $10^6$ $M_\odot$ for globular clusters in this density and concentration range. For the Model 4.1 runs we find two of the collisions had the third star nominally bound after merger, but in orbits wide enough that the third star should be stripped from the pulsars on time scales of $O(10^8)$ years. For Model 4.2 we find one case where the third star (another NS) should be bound in a tight, $a \sim 0.2\,AU$, orbit; one MS–WD merger with a bound ($a \sim 0.5\,AU$) companion, and one WD–WD merger AIC candidate with a bound WD at $a \sim 0.5\,AU$. Thus we conclude $\sim 75\%$ of MSPs in clusters with $n_0 \sim 10^5$ should be single, or have a low mass stellar core remnant in a very tight orbit, such as PSR B1744-24 in Ter5, or PSR B0021-21I,J in 47Tuc.

For Models 4.3.1,2,3, for which the IMF slope $x_* = 1.00$, we have 9 NS collisions. Inferred $\Gamma_{MSP-coll} = 72$, or 36 observable MSPs given a beaming factor of 0.5. Model 4.3 is comparable to Ter5, in which a number of unresolved steep spectrum radio sources are observed (Fruchter & Goss 1990). Of the 9 NS collisions, 5 led to the third star unbound to the merger, one involved a MS colliding with a MS–NS contact binary and probable triple merger, three led to the third star bound in a tight orbit. In addition, there were 4 WD–WD collisions three of which could in principle lead to AIC. There were 2 WD–MS collisions in which the third star was a NS and was inferred to remain bound to the merged star on a tight orbit. For a total $\Gamma_{MSP} = 120$! It is worth noting that one of the NS collisions involved an initial WD–NS binary with $a = 0.1\,AU$, encountering a field WD which collided with the NS, leaving the original WD companion on a $0.4\,AU$ orbit.

For our densest model, Model 5, we have 13 NS collisions. From runs 5.0.1 and 5.0.4, the inferred $\Gamma_{MSP-coll} = 31$ or 15-16 observable MSPs given our assumed beaming factor and MSP lifetime. By comparison, run 5.0.2 where we assumed too large a range in $a$ and $m_{1,2}$ for the binaries surviving in the core, leads to much larger inferred $\Gamma_{MSP-coll}$ ($= 1100$). In addition we had 6 WD–WD collisions that might lead to AIC, and 4 WD–MS collisions in which the third star was a NS inferred to be bound in an orbit tight enough for subsequent mass–transfer. Two of the WD–MS + NS collisions would probably have led to the NS colliding promptly with the WD–MS merged star, the other two led to the NS bound in an orbit of $0.1 - 0.5\,AU$, two of the WD collisions were in model 5.0.2 and are discarded for the MSP calculation. This gives a total $\Gamma_{MSP} \approx 62$ or 31 observable MSPs.

### 4.7.2. White Dwarfs

In all the Model 3 runs, 7 MS–WD mergers were observed during our simulations. All came from wide ($a \sim 0.3 - 3\,AU$) MS–WD binaries, with the stars colliding during a resonant encounter, leaving the third star unbound in our approximation. The MSs had masses of $0.5 - 0.7\,\mathrm{M}_{\odot}$, while the WDs were $0.5 - 0.9\,\mathrm{M}_{\odot}$. Averaging over the models discarding the low initial concentration runs which produced no collision, we infer $\Gamma_{CV-coll} = 79$, for cluster with $n_0 \sim 10^4$, $M_T \sim 10^6$. Interestingly, the largest contribution to $\Gamma$ is from the lower mass and lower initial concentration models, suggesting care must be taken in inferring total $\Gamma$ for the more concentrated models from the high mass binary runs.

There were 3 WD–MS collisions in Models 4.1, giving a $\Gamma_{CV-coll} = 116$, however, using just Model 4.1.1 we infer $\Gamma_{CV-coll} = 330$, suggesting that in fact the high binary mass models may be missing collisions of low mass WDs. For Model 4.2.1, we have 27 WD collisions, 4 of which have been assumed to lead to MSPs, leaving a total $\Gamma_{CV-coll} = 100$, Model 4.2 is twice as dense as Model 4.1 but only half as massive, suggesting our estimates for CVs formed through the T3 channel are robust. In Models 4.3 there were 14 WD collisions, 5 of which were assumed to lead to an MSP, leaving a $\Gamma_{CV-coll} = 1590$.

In Models 5, we have 16 WD collisions, 10 of which were assumed to lead to MSPs, from Models 5.0.1 and 5.0.4 we infer $\Gamma_{CV-coll} = 776$, while from Model 5.0.2 we infer $\Gamma_{CV-coll} = 6600$! It is likely the real production rate is intermediate, with the massive binary runs undercounting the CV formation rates, while run 5.0.2 extends to large $a$ than are likely to survive in a cluster of such a high density, even allowing for halo binaries returning to the core. A further problem is caused by the fact that $\tau_{CV} \gg T$ for Model 5, and many of the observable CVs would have been formed before core–collapse, while the core was at a lower density.

### 4.7.3. Main Sequence Stars

There were 8 MS–MS collisions in the model 3 simulations. Seven of those eight involved stars whose combined mass exceeded the turnoff and would lead presumably lead to a BS formed. The remaining collision, in model 3.3.3, was between a $0.4\,\mathrm{M}_{\odot}$ and a $0.3\,\mathrm{M}_{\odot}$ star and may not have produced a BS. We infer $\Gamma_{BS} = 15$, averaging over the runs. For comparison, we infer $O(150)$ low mass stars ($m_1 + m_2 \le 0.7\,\mathrm{M}_{\odot}$) have collided in the last $10^{10}$ years per $10^6\,\mathrm{M}_{\odot}$ for moderate density clusters.

The MS–MS collisions in Models 4.1 and 4.2 are not statistically significant. For Model 4.2.1 we have 12 MS–MS collisions, one of which was assumed to lead to a MSP. Of the remaining 11 collisions 2 involved a giant or a sub–giant, for a total $\Gamma_{MS-MS} = 80$. It should be noted that the mass range for the binaries in Model 4.2.1 is too high to cover the lower end of the blue straggler range, so this should be considered an estimate for the brighter blue straggler population only.

We can also estimate $\Gamma_{MS-MS}$ from Models 4.3.1,3 and 5.0.2 which include binaries in the appropriate range in mass. For Model 4.3.1,3 we have 5 MS–MS collisions with no NS bound, for a $\Gamma_{MS-MS} = 1470$. For Model 5.0.2 we have 5 MS–MS collisions (one involving a (sub)giant), for a $\Gamma_{MS-MS} = 27500$! Again, the estimate for Model 5 is not reliable because of the excessively large range in $a$ and the fact that $\tau_{BS} \gg T$ and most of the MS–MS collisions should have taken place prior to core–collapse.

# 5. Conclusion

Binary–single star interactions (T3) are a major production channel for recycled pulsars in globular clusters, and binary encounters also contribute significantly to production of cataclysmic variables and blue stragglers in globular clusters. Exchange production (E) also appears to be a significant channel for MSP formation in intermediate density clusters. Other possible channels include single star tidal interaction (T2), binary–binary interactions (T4) and AIC (see Phinney & Kulkarni).

## *5.1. Pulsars in Globular Clusters*

There are 32 currently known pulsars in galactic globular clusters. Eight are in M15, another ten are in 47Tuc. The detection of pulsars in globular clusters is strongly biased by selection effects: including which globular clusters can be reached by the biggest radio observatories, notably Arecibo; different dispersion measure towards different clusters, and simply the proximity of the different clusters to our solar system. A major selection problem is due to the variable pulsar spin period observed in short period binary pulsars, which substantially reduces search sensitivities to short period binary pulsars and biases the observed single to binary pulsar ratio (see Phinney & Kulkarni 1994, and references therein). A more subtle problem is that whether considered by individual globular cluster, groups of similar clusters, or in total, the statistical sample of pulsars in globular clusters is poor. Our modeling of globular cluster pulsars is intrinsically limited by this poor statistical population and some of the pulsars observed may be individually "unique" rather than representative of an underlying population.

Additional confounding factors include the uncertainty in the spin–up mechanism for recycled pulsars, which may lead to systematic differences in pulsar populations in clusters of different concentration. Both the question of the relative importance of classical LMXBs vs some rapid, non–observed channel for neutron star spin–up, and the question of AIC, may cause population gradients between clusters of different density and concentration. Further problems are caused by the lack of characteristic ages for many cluster MSPs and the contamination of the measured ages by pulsar acceleration (Phinney 1992).

47Tuc (NGC 104) is the most interesting of the globular clusters containing observed MSPs. All 10 observed MSPs are short spin period, at least five are thought to be single, two are in near contact binaries and one is a short period binary. From our models we would conclude that binary–single star encounters can account for the 47Tuc population if the cluster is intermediate between our Models 4.1,2 and 4.3. That is, a IMF with a relatively flat slope, leaving $\sim 2\%$ NS remnants by mass, $\sim 20\%$ of which remain bound to the cluster, and a concentrated population of $a \sim 0.1\, AU$ massive binaries will produce the observed number of pulsars, with the majority single MSPs.

We predict a large number of MSPs should be present in Ter5. Only one is observed, but VLA imaging suggests many more are present, not yet detected by timing. PSR 1744-24A position outside the core of Ter5, with a low mass, accreting companion is consistent with a binary–single encounter involving a collision and modest recoil (Sigurdsson 1991, Phinney & Sigurdsson 1991). The number of MSPs detectable in principle in Ter5 should be 10-30, if its binary and remnant population is comparable to that of other globulars.

For M15 (NGC 7078), we infer too many observable MSPs from our model. Comparing with run 5.0.2, and the final state of runs 4.2,3, we can see that our canonical estimates for $f_b$ cannot hold in core–collapsed clusters, too many binaries are destroyed or ejected before the cluster can reach its present density to sustain the interaction rate inferred. Allowing for a reduced $f_b$, and the expected contribution to the MSP production by neutron star–single star collision (Krolik *et al.* 1984, Phinney & Kulkarni 1994), our model is consistent with the number of MSPs observed. We require that $\sim 2/3$ of the hard, massive binaries in M15 have been disrupted or ejected from the core by past interactions. Certainly the spectacular success of binary–single star encounter models in explaining PSR B2127+11C suggests that T3 production is significant in M15 (Phinney & Sigurdsson 1991).

Cluster NGC 6624 is a high density cluster with relatively low dispersion and only $\sim 1/10$ of the mass of M15. It contains two observed single pulsars, one short spin period one long spin period. This is consistent with the population in M15, and reinforces the suggestion that the long spin period pulsars in these dense clusters may have formed by T2, and the short spin period pulsars formed by T3 (or related T4/AIC processes). This would suggest that the single slow pulsar in NGC 6440 may have formed by T2.

PSR B1821-24A in M28 (NGC 6626) is the original globular cluster pulsar. M28 is intermediate between our Model 3.2 and Model 4.2 suggesting we should indeed expect to see $O(1)$ MSP in the cluster, either a LMBP formed by exchange or a 47Tuc class single pulsar formed by T3. PSR B1802-07A in NGC 6539 is a comparable case, but in a short period eccentric binary. Most likely PSR B1802-07A formed by T3 or T4, with the present companion the remnant of the colliding (turnoff mass) star, with the secondary envelope ejected early in the spin–up process and circularization of the binary terminated as the remnant core detached. Alternatively, it could be the spectator (white dwarf) star in a T3/T4 collision, the colliding (main–sequence) star having been entirely disrupted. This is somewhat less likely, because of the very tight orbit of the secondary, and would have required the current secondary undergo a common–envelope evolution phase as the colliding star remnant was dissipated, in order to bring the secondary to its current orbit.

M5 (NGC 5904) contains two pulsars, PSR B1516+02A,B, one single, one in a 7 day eccentric binary. With a density of $\approx 10^4\,\mathrm{M}_\odot\,\mathrm{pc}^{-3}$ and moderate dispersion, M5 is closest to our Model 3.2, where we would predict one MSP per several clusters. M5A is an unusually long characteristic age ($2 \times 10^9$ years), which partly accounts for the number observed. Both pulsars are consistent with a T3/T4 collision, with M5B similar to PSR B1802-07A, and a possible detached remnant of a star that underwent a grazing collision with the neutron star. If M5 has a high core binary fraction, the presence of two pulsars is readily explained by the enchanced rate for T4 production over T3.

PSR B1620-26 and PSR B1310+18 in M4 (NGC 6121) and M53 (NGC 5024) are most certainly exchange binaries as discussed earlier. PSR B1908+00A, a single pulsar in NGC 6760, and the two pulsars in M13 (NGC 6205), single PSR B1639+36A and binary PSR B1639+36B fall into the same category as the M5 pulsars, and may also be accounted for by a combination of T3 and T4 collisions given a Salpeter IMF and a high binary fraction. The lower dispersion of these clusters enhances the cross–section for T3 (and T4) processes relative to our models and partially accounts for the higher MSP production rate inferred. We will be exploring T3 processes in low mass, low dispersion clusters further using our technique.

We do not find it necessary to invoke AIC to account for any globular cluster MSP formation, although it is by no means excluded as a formation channel. The relative fraction of NS spin–up to AIC depends on $f_r$ (which we assume to be constant but may vary according to cluster concentration and even metallicity), compared to the ratio of HWDs that may undergo AIC to NS formed given the assumed IMF. The latter factor is very poorly constrained observationally and theoretically, as is the actual mass and accretion rate necessary for AIC. Upon undergoing AIC, a binary should go from $e = 0$ to $e \sim 0.1$ due to the sudden loss of mass from neutrino flux, and this might be a signature of AIC, with M5B and possible PSR B1802-07A being possible candidates; however, if the secondary does not become detached at AIC but continues mass–transfer, then the binary will recircularize and continue evolving. Detailed modeling of AIC scenarios is necessary to resolve this issue. If AIC does not occur, then the WD–WD collisions observed, for which the total mass exceeds the Chandrasekhar mass, should lead to detonation and an observable supernova. We find WD–WD collisions occurring at the rate of $R_A \sim 5 \times 10^{-10}\,\mathrm{y}^{-1}$ for $n_0 \sim 10^5\,\mathrm{pc}^{-3}$, $R_A \sim 3 \times 10^{-8}\,\mathrm{y}^{-1}$ for flat IMF, $n_0 \sim 3 \times 10^5\,\mathrm{pc}^{-3}$ Model 4.3, and $R_A \sim 5 \times 10^{-7}\,\mathrm{y}^{-1}$ for our Model 5. Averaged over the galaxy, we infer a globular cluster supernova rate of $R_{A-SN} \sim 3 - 5 \times 10^{-6}\,\mathrm{y}^{-1}$, dominated by core–collapsed clusters, possibly a factor of ten higher if the IMF is flat or T4 processes contribute strongly ($f_b$ large). This compares with a total galactic supernova rate estimated at $\sim 10^{-2}\,\mathrm{y}^{-1}$. We are unlikely to observe an extragalactic supernova in a globular cluster in the near future but may see one per century at current detection rates.

Our simulations show a few LMBPs should be formed in intermediate density clusters with companions with $O(0.1\,AU)$ eccentric orbits, formed by T3 (or T4) collisions, the third star remaining bound to the merged remnant after the collision. M5B is a possible example of such an object, we expect more to be found in slightly wider, more eccentric orbits. The extent to which the secondary orbit would be circularized during the spin–up of the MSP and the ejection of the bulk of the collided star is a largely open question and one of current theoretical interest.

We do not consider the process of spin–up here, and hence cannot make strong statements about expected number of X–ray sources, Thorne–Żytkow objects (Thorne & Żytkow 1977) or "shrouded" neutron stars (Krolik *et al.* 1984, Tavani 1991).

## 5.2. *Blue Stragglers and Cataclysmic Variables*

From our models, we can make some qualified statements about the production of CVs and BSs through the T3 channel. Both CVs and BSs may be expected to be currently formed in globular clusters through primordial binary evolution, as well as through interactions. T3 production, as well as T2 and T4 channels, will add to this background rate. In core–collapsed clusters densities are high enough that binary interactions and exchanges have significantly decreased the production rate from primordial, isolated binaries.

For CVs, we predict T3 production, relative to the MSP production, is relatively more efficient at low to intermediate densities than for high density clusters, with $N_{CV-T3}/N_{MSP-T3}$ decreasing from $\sim 200$ for $n_0 \sim 10^4$ to $\sim 20$ for core–collapsed clusters like M15. Comparing the number of bright CVs to the number of MSPs observed in different density clusters should constrain the relative importance of different production channels, and hence the mass function and binary fraction.

Binary interactions have been suggested for blue straggler formation in globular clusters. Leonard has considered this scenario in considerable detail (Leonard 1989, Leonard and Fahlman 1991). We find that T3 production can account for $10-100$ BS per $10^6\,M_\odot$ in intermediate density clusters. At high densities, the MS collisions rate is dominated by encounters with degenerate stars and we'd expect proportionately fewer BS. Artificially forcing a population of MS binaries into the core of high density clusters can lead to large numbers of BS formed, as seen in run 5.0.2, but this is not representative of real clusters. T2 interactions should be proportionately more important for BS formation in the high density clusters (diStefano & Rappaport 1994), and T4 interactions may dominate for low density clusters (Leonard 1989). The ratio of BSs to CVs and MSPs should further constrain the IMF and binary fraction in different clusters.

## 5.3. Other Factors and Future Problems

It is important to remember that half the mass of a globular cluster is beyond the half–mass radius, we cannot model a globular cluster by its core. The second lesson is that, unfortunately, systematic uncertainties in globular cluster parameters are still a major source of error in modeling cluster processes. In particular, the direct constraints on the global IMF and the global, total primordial binary fraction are too weak. Our models do allow some indirect constraints to be made on globular cluster properties, given our associated assumptions, and, perhaps surprisingly, the canonical parameters assumed for many cluster properties lead naturally to many of the more exotic observed properties. That is, a Salpeter IMF with $\sim 2\%$ NS remnants by mass and a modest $f_r \approx 0.2$ is adequate to account for the MSPs observed through T3 and T4 channel collisions, provided the total, global primordial binary fraction in globular clusters is comparable to that in the galaxy.

There is some degeneracy between $f_b$ in the $a \sim 0.1 - 1.0\,AU$ range, and the $N_{NS}(IMF) \times f_r$, but this is constrained by the number of MSPs in different concentration globular clusters. We infer the $N_{NS}(IMF)$ is not much larger than expected from the Salpeter IMF with our choice of progenitor mass $\mapsto$ remnant mass, or we would expect proportionately more MSPs in 47Tuc class clusters. We do require that $f_r$ is not too small, implying that neutron stars are born with a velocity distribution that has a significant tail at low velocities (at least for low metallicity progenitors), or AIC is very efficient. The value of $f_r$ favored ($\sim 0.2$) is comfortably within the theoretically permitted range (Hut & Verbunt 1983) and consistent with recent models of NS retention in zero age globular clusters (Sigurdsson & Hernquist [in preparation]). We do favor high $f_b$, at least 10% of stars in globular clusters should be $\sim 0.1 - 1.0\,AU$ binaries, this is strongly constrained by the excess of MSPs in the lowest density clusters, where a large $f_b$ may enhance T4 interactions relative to T3 sufficiently to account for the MSPs observed. Alternatively, the binary semi–major axis distribution is peaked around $0.2\,AU$, as suggested by Trimble (1976)

The effectiveness of binary–binary interactions needs to be estimated to evaluate the validity of the assumptions used in our calculations. Binary–binary interactions are significant if

$$\frac{\tilde{\sigma}_4 f_b}{\tilde{\sigma}_3 (1 - f_b)} \sim 1. \tag{5.1}$$

The total cross-section for binary–binary interactions is approximately four times that of binary–single star encounters. The collision cross–section during binary–binary encounters is dominated by resonances, during which, typically, the lightest star is promptly ejected and the remaining three stars undergo a "democratic" resonance. The distribution of close approaches during such encounters is dependent only on the angular momentum and energy of the triple (Valtonen 1988), not whether the triple formed from a binary–binary encounter or a binary–single encounter. As the fractional cross–section for destruction of one of the binaries is high (Mikkola 1975, Hut *et al.* 1992), binary–binary interactions are a self–limiting process, if the core $f_b$ is high, binaries are rapidly destroyed until the interaction rate is small. Mass segregation is effective at increasing the relative core $f_b$ in concentrated clusters, in low concentration clusters relaxation time scales are long and the total core mass is a large fraction of the cluster mass, requiring that a high partial density of low mass, single stars in the core.

Calculating the full differential cross-section for binary–binary encounters for a realistic range of masses and semi–major axis will be necessary to determine the actual efficiency of binary–binary encounters, and getting definitive estimates of the different interaction rates. In the meantime, the detection of blue–stragglers in low density clusters (Nemec and Cohen 1989, Yan & Mateo 1994, Bolte 1992) may be taken to provide a measure of the background rate, due perhaps to merger of close primordial binaries during evolution (Mateo *et al.* 1990). The encounter rate observed, and the parameters of the colliding systems involving neutron stars are consistent with the ratio and distribution of single and binary pulsars observed in the Galactic globular system. In particular, the ratio of single to binary pulsars is accounted for, and the presence of binary pulsars M15C and Ter5A outside their cluster cores is explained. As found previously (Phinney and Sigurdsson 1991) a number of binary neutron stars were found to be ejected from the cluster after colliding, and of order 20 short period binary pulsars with a neutron star or white dwarf companion might be expected in the Galaxy, having been ejected from core collapsed globular clusters in the last $10^9$ years(Phinney and Sigurdsson 1991).

The evolution of a complete range of binaries in a time evolving cluster model, from zero age through collapse, is necessary to ultimately determine the correct parameters. Such a calculation should allow for stellar evolution and binary orbital evolution through gravitational radiation. The heating rates calculated with these models and in Sigurdsson and Phinney (1993) should contribute to the development of a self–consistent cluster evolution model. The inclusion of binary–binary encounters is necessary if $f_b \sim 0.5$, and it would be desirable to include encounters with hierarchical trinaries. Some progress is being made towards systematically calculating such encounters (Hut 1990). At a later stage we expect to make calculations in a time varying background. By using the estimated energy release during the evolution, and iterating the calculation of the evolution of the cluster collapse, and using SPH codes to model the outcome of stellar collisions (Davies *et al.* 1994), we hope to eventually produce partially self–consistent models of cluster evolution. We also plan to investigate the effects of a population of low mass black holes on the cluster dynamics (Grindlay 1993, Kulkarni *et al.* 1993, Sigurdsson & Hernquist 1993).

It is possible to account for the pulsars observed in the Galactic globular clusters through the interactions of binaries and neutron stars, assuming some very reasonable values of the globular cluster parameters.

We would like to thank D. Chernoff for helpful discussion and comments. This research was supported in part by NASA grants NAGW–2144 and NAGW–2422, the NSF under Grants AST 90–18526 and ASC 93-18185, and an ASF Thor Thors fellowship.

# Captions

Table 1

Table of mass–groups for the models run. $x_*$ is the exponent of the initial mass function, the second column shows the index of mass groups. $m_{hi}$ and $m_{lo}$ are the upper and lower boundaries on each mass group, respectively, while $\bar{m}$ is the mean mass of that mass group. $f_m$ is the fraction of the total mass of the cluster in that mass group, $f_n$ is the number fraction of that mass group in the cluster and $f_L$ is the fraction of that mass group that is luminous. Model 4.3 was run with $x_* = 1.00$ but only had eight mass groups; the lowest mass group covered the range covered by mass groups 1–3 in this table, the other mass groups were the same. The mean stellar mass for $x_* = 1.00, 1.35, 1.50$ was $0.224, 0.241, 0.297\,M_\odot$ respectively.

Table 2

The parameters of the models used in the calculations. $x_*$ is the exponent of the initial mass function, as before, $W_0$ is the ratio of the depth of the potential to the mean core dispersion, as described in the text. $r_c$ is the core radius given in terms of the scale radius, $r_0$. $r_t$ is the tidal radius of the cluster, $r_0$ is the mean "King radius" as defined in the text, $n_0$ is the core density, $\bar{m}_c$ is the mean stellar mass in the core, $\bar{\sigma}(0)$ is the mean core dispersion in km s$^{-1}$, and $M_T$ is the total mass of the cluster.

Table 3

The parameters of the individual model runs. The runs are arbitrarily labeled, and the number of binaries, $N_b$, in each set of runs is given. Sets of runs for which the number of binaries is not a multiple of 50 were prematurely terminated by computer crashes. The third column gives the mass group index according to whose radial distribution the binaries were initially placed. We expect the relaxed distribution of the binaries to be concentrated (neglecting encounter recoil), with the heavier binaries more concentrated than the most concentrated single star mass group. For Columns four and five show the integration time and the "super–orbit" scale factor respectively. Columns six and seven show the range in initial binary semi–major axis in AU. The range was generally chosen so that the maximum size binary was marginally soft. Exceptions are a couple of runs with very wide binaries to check ionization rates, and the high initial concentration runs for Model 5 where wide binaries are not expected to be found at all. The minimum semi–major axis was generally chosen such that a binary with $a = a_{min}$ had an encounter probability $\ll 1/N_b$ in the core, over the integration time. Columns eight and nine show the minimum mass imposed on the stars in the binary in $M_\odot$. The last column shows the binary weight, $f_w$, the fraction of the total number of binaries that have the members with masses in the range imposed, assuming the binaries drawn independently from the IMF. The weight does not include the restriction in semi–major axis which should be factored separately into $f_b$.

Table 4.

A summary of the outcome of individual runs. The second column shows the number of ionized binaries. The third and fourth columns show the number of exchanges and collisions. Columns 5–7 show respectively the number of collisions involving neutron stars, white dwarfs, and main–sequence stars only, as discussed in section 4.

Figure 1

Concentration $c = \log(r_t/r_c)$ vs $W_0$ for multi–mass models with different $x_*$ compared with the single mass concentration. The multi–mass models span a smaller concentration range as a function of $W_0$, forcing a choice of larger $W_0$ to fit observed concentrations.

Figure 2a

Relative orbit averaged integrated encounter probabilities for different mass groups (shown by index, 1 being the lowest mass group, 10 the highest), for a $1\,AU$ binary in Model 1.

Figure 2b

Relative orbit averaged integrated encounter probabilities for different mass groups for a $1\,AU$ binary in Model 3.4.

Figure 2c

Relative orbit averaged integrated encounter probabilities for different mass groups for a $1\,AU$ binary in Model 4.2.

Figure 2d

Relative orbit averaged integrated encounter probabilities for different mass groups for a $1\,AU$ binary in Model 5.

Figure 3

The fraction of time a star ejected from a cluster core can be observed, in projection, at different speeds, $v_p$. The $1.0\,\mathrm{M}_\odot$ star was ejected from the center of the cluster core with initial speed $0.9\sqrt{2W_0}$. Two curves are show for each model ($x_* = 1.35, W_0 = 6, 9, 12$), one averaging $v_p$ for the first $10^8$ years after ejection, the other showing the average over $10^9$ years of orbit evolution. The intrinsic dispersion, $\sigma$ ($\bar{\sigma}$ in text), is not equal to the projected turnoff mass group disperson, which is the observed dispersion, the latter is $\sim 20\%$ smaller.

Figure 4a

The initial (dashed line) and final (solid line) radial binary distribution in run 2.0.1, showing the relaxation of the binary distribution. The plot shows the number of binaries per octave in radius.

Figure 4b

Radial binary distribution in run 2.0.2 showing the relaxation of massive binaries to the core of the cluster.

Figure 4c

The radial binary distribution for a set of Model 3 runs, showing relaxation and the start of ejection of binaries due to encounters.

Figure 4d

The radial binary distribution in runs 4.1.2 and 4.1.3 averaged together. The plots shows the pronounced ejection of binaries to the half–mass radius in denser clusters.

Figure 4e

The radial binary distribution in Model 4.3, showing the strong mass–segregation of this flat IMF cluster, pronounced ejection to the half–mass radius, and the start of ejection of binaries from the cluster.

Figure 4f

The radial binary distribution for the set of Model 5 runs showing again the strong segregation to the core of the massive binaries, ejection to the halo and ejection from the cluster.

Figure 5a

The initial (dashed line) and final (solid line) orbital period (in days) distribution for Model 1.0.1. Binaries that ionized (or collided leaving a single star) are shown at $\log(P) = -3$. In this low density cluster there is little change in the binary period distribution, mostly some modest ionization of the widest binaries. It should be noted that the binary period is a function of both $a$, and the binary mass, drawn from the IMF with the constraint from Table 3.

Figure 5b

The initial and final orbital period distribution for Model 2.1.2 showing the ionization of the longer period binaries in moderate density clusters.

Figure 5c

The initial and final orbital period distribution for Model 3.2.4 showing some hardening of binaries and ionization of the longest period binaries.

Figure 5d

The initial and final binary period distribution in Model 4.2.1. In this case only about half the binaries at $\log(P) = -3$ are ionized, the rest being collisions. Interestingly the short period binary population is not heavily modified showing the collisions effectively draw from the wider interacting binaries. On average a binary is hardened to replace any short period binary that collides.

Figure 5e

The initial and final binary period distribution in Model 5.0.1. Only three binaries were ionized in this run, the bulk at $\log(P) = -3$ being due to collisions. A substantial fraction of the collided binaries are effectively drawn from the longer period population with hardening replenishing the collided short period binaries.

# Tables

**Table 1.**

Multi–mass models

| $x_*$ | Mass group | $m_{hi}$ | $m_{lo}$ | $\bar{m}$ | $f_m$ | $f_n$ | $f_L$ |
|---|---|---|---|---|---|---|---|
| 1.35 | 1 | 0.157 | 0.100 | 0.1235 | 0.2339 | 0.4571 | 1.000 |
| | 2 | 0.200 | 0.157 | 0.1761 | 0.1105 | 0.1514 | 1.000 |
| | 3 | 0.250 | 0.200 | 0.2228 | 0.0975 | 0.1056 | 1.000 |
| | 4 | 0.310 | 0.250 | 0.2779 | 0.0824 | 0.0716 | 1.000 |
| | 5 | 0.390 | 0.310 | 0.3460 | 0.0836 | 0.0583 | 1.000 |
| | 6 | 0.500 | 0.390 | 0.4396 | 0.0836 | 0.0459 | 1.000 |
| | 7 | 0.630 | 0.500 | 0.5668 | 0.1343 | 0.0572 | 0.526 |
| | 8 | 0.800 | 0.630 | 0.7042 | 0.1155 | 0.0396 | 0.588 |
| | 9 | 1.250 | 0.800 | 0.9659 | 0.0428 | 0.0107 | 0.000 |
| | 10 | 1.570 | 1.250 | 1.3634 | 0.0158 | 0.0028 | 0.000 |
| 1.00 | 1 | 0.157 | 0.100 | 0.1242 | 0.1539 | 0.3681 | 1.000 |
| | 2 | 0.200 | 0.157 | 0.1764 | 0.0821 | 0.1382 | 1.000 |
| | 3 | 0.250 | 0.200 | 0.2232 | 0.0787 | 0.1047 | 1.000 |
| | 4 | 0.310 | 0.250 | 0.2782 | 0.0718 | 0.0767 | 1.000 |
| | 5 | 0.390 | 0.310 | 0.3465 | 0.0787 | 0.0674 | 1.000 |
| | 6 | 0.500 | 0.390 | 0.4404 | 0.0855 | 0.0577 | 1.000 |
| | 7 | 0.630 | 0.500 | 0.5686 | 0.1647 | 0.0861 | 0.477 |
| | 8 | 0.800 | 0.630 | 0.7055 | 0.1577 | 0.0664 | 0.520 |
| | 9 | 1.250 | 0.800 | 0.9745 | 0.0835 | 0.0255 | 0.000 |
| | 10 | 1.570 | 1.250 | 1.3697 | 0.0436 | 0.0094 | 0.000 |
| 1.50 | 1 | 0.157 | 0.100 | 0.1231 | 0.2703 | 0.4919 | 1.000 |
| | 2 | 0.200 | 0.157 | 0.1760 | 0.1211 | 0.1543 | 1.000 |
| | 3 | 0.250 | 0.200 | 0.2227 | 0.1032 | 0.1039 | 1.000 |
| | 4 | 0.310 | 0.250 | 0.2777 | 0.0844 | 0.0681 | 1.000 |
| | 5 | 0.390 | 0.310 | 0.3458 | 0.0828 | 0.0537 | 1.000 |
| | 6 | 0.500 | 0.390 | 0.4392 | 0.0799 | 0.0408 | 1.000 |
| | 7 | 0.630 | 0.500 | 0.5660 | 0.1191 | 0.0472 | 0.547 |
| | 8 | 0.800 | 0.630 | 0.7037 | 0.0981 | 0.0313 | 0.616 |
| | 9 | 1.250 | 0.800 | 0.9622 | 0.0311 | 0.0072 | 0.000 |
| | 10 | 1.570 | 1.250 | 1.361 | 0.0099 | 0.0016 | 0.000 |

**Table 2.**

Cluster models

| Model | $x_*$ | $W_0$ | $r_c/r_0$ | $r_t/r_c$ | $r_0$/pc | $r_h/r_c$ | $n_0$ $/pc^3$ | $\bar{m}_c/\mathrm{M}_\odot$ | $\bar{\sigma}(0)$ | $M_T/\mathrm{M}_\odot$ |
|---|---|---|---|---|---|---|---|---|---|---|
| 1.0 | 1.50 | 4 | 0.69 | 12.1 | 5.7 | 2.78 | $2 \times 10^2$ | 0.38 | 4 | $1.4 \times 10^5$ |
| 2.0 | 1.00 | 6 | 0.95 | 17.1 | 3.1 | 3.47 | $1 \times 10^3$ | 0.63 | 6 | $3.7 \times 10^5$ |
| 2.1 | 1.00 | 10 | 1.30 | 39.7 | 1.95 | 7.77 | $4 \times 10^3$ | 0.89 | 9 | $2.0 \times 10^6$ |
| 3.0 | 1.35 | 12 | 1.25 | 58.6 | 1.25 | 11.8 | $8 \times 10^3$ | 0.84 | 7 | $1.2 \times 10^6$ |
| 3.1 | 1.35 | 10 | 1.10 | 42.6 | 1.21 | 8.34 | $1 \times 10^4$ | 0.73 | 8 | $9.4 \times 10^5$ |
| 3.2 | 1.35 | 12 | 1.25 | 58.6 | 1.13 | 11.8 | $1 \times 10^4$ | 0.84 | 8 | $1.6 \times 10^6$ |
| 3.3 | 1.35 | 9 | 1.00 | 36.5 | 1.41 | 7.11 | $1 \times 10^4$ | 0.68 | 9 | $1.0 \times 10^6$ |
| 3.4 | 1.35 | 6 | 0.82 | 19.4 | 1.52 | 3.96 | $2 \times 10^4$ | 0.52 | 12 | $7.4 \times 10^5$ |
| 4.0 | 1.35 | 12 | 1.25 | 58.6 | 0.65 | 11.8 | $8 \times 10^4$ | 0.84 | 13 | $2.4 \times 10^6$ |
| 4.1 | 1.35 | 12 | 1.25 | 58.6 | 0.62 | 11.8 | $1 \times 10^5$ | 0.84 | 14 | $2.6 \times 10^6$ |
| 4.2 | 1.35 | 12 | 1.25 | 58.6 | 0.38 | 11.8 | $2 \times 10^5$ | 0.84 | 12 | $1.2 \times 10^6$ |
| 4.3 | 1.00 | 12 | 1.57 | 54.3 | 0.35 | 11.2 | $3 \times 10^5$ | 1.01 | 15 | $1.9 \times 10^6$ |
| 5.0 | 1.35 | 18 | 1.98 | 133 | 0.11 | 28.9 | $3 \times 10^6$ | 1.16 | 15 | $2.6 \times 10^6$ |

**Table 3a.**

Run parameters

| run | $N_b$ | In. conc. | $T$ | $t_n$ | $a_{max}$ | $a_{min}$ | $M_{1_m}$ | $M_{2_m}$ | $f_w$ |
|---|---|---|---|---|---|---|---|---|---|
| 1.0.1 | 2500 | 8 | $10^{10}$ | 20 | $10^3$ | 10.0 | 0.15 | 0.00 | $5.1 \times 10^{-1}$ |
| 1.0.2 | 2500 | 4 | $1.5 \times 10^{10}$ | 20 | $10^3$ | 10.0 | 0.50 | 0.35 | $1.1 \times 10^{-2}$ |
| 2.0.1 | 1000 | 8 | $10^{10}$ | 40 | 10.0 | 0.10 | 0.35 | 0.20 | $1.2 \times 10^{-1}$ |
| 2.0.2 | 1000 | 8 | $10^{10}$ | 50 | 50.0 | 0.50 | 0.45 | 0.25 | $7.3 \times 10^{-2}$ |
| 2.1.1 | 800 | 8 | $10^{10}$ | 100 | 5.0 | 0.05 | 0.25 | 0.00 | $2.9 \times 10^{-1}$ |
| 2.1.2 | 500 | 8 | $10^{10}$ | 100 | 100.0 | 5.0 | 0.50 | 0.35 | $1.7 \times 10^{-2}$ |
| 3.0.1 | 1000 | 8 | $10^{10}$ | 5 | 20.0 | 0.20 | 0.50 | 0.35 | $1.7 \times 10^{-2}$ |
| 3.1.1 | 1000 | 6 | $1.5 \times 10^{10}$ | 5 | 10.0 | 0.10 | 0.50 | 0.35 | $1.7 \times 10^{-2}$ |
| 3.1.2 | 446 | 8 | $1.5 \times 10^{10}$ | 5 | 30.0 | 0.03 | 0.50 | 0.35 | $1.7 \times 10^{-2}$ |
| 3.2.1 | 1000 | 8 | $10^{10}$ | 5 | 10.0 | 0.10 | 0.50 | 0.35 | $1.7 \times 10^{-2}$ |
| 3.2.2 | 235 | 10 | $10^{10}$ | 5 | 10.0 | 0.10 | 0.70 | 0.50 | $5.9 \times 10^{-3}$ |
| 3.2.3 | 1000 | 8 | $10^{10}$ | 5 | 10.0 | 0.10 | 0.70 | 0.50 | $5.9 \times 10^{-3}$ |
| 3.2.4 | 1000 | 10 | $10^{10}$ | 5 | 20.0 | 0.20 | 0.70 | 0.50 | $5.9 \times 10^{-3}$ |
| 3.2.5 | 1000 | 8 | $10^{10}$ | 5 | 20.0 | 0.20 | 0.50 | 0.20 | $4.3 \times 10^{-2}$ |
| 3.2.6 | 1000 | 10 | $10^{10}$ | 5 | 10.0 | 0.10 | 0.63 | 0.40 | $8.3 \times 10^{-3}$ |
| 3.3.1 | 250 | 6 | $10^{10}$ | 20 | 10.0 | 0.05 | 0.50 | 0.35 | $1.7 \times 10^{-2}$ |
| 3.3.2 | 250 | 6 | $10^{10}$ | 100 | 10.0 | 0.05 | 0.20 | 0.00 | $3.9 \times 10^{-1}$ |
| 3.3.3 | 250 | 10 | $10^{10}$ | 100 | 10.0 | 0.05 | 0.00 | 0.00 | 1.0 |
| 3.3.4 | 250 | 6 | $10^{10}$ | 50 | 10.0 | 0.05 | 0.50 | 0.35 | $1.7 \times 10^{-2}$ |
| 3.4.1 | 500 | 6 | $10^{10}$ | 20 | 20.0 | 0.10 | 0.50 | 0.35 | $1.7 \times 10^{-2}$ |

**Table 3b.**

Run parameters

| run | $N_b$ | In. conc. | $T$ | $t_n$ | $a_{max}$ | $a_{min}$ | $M_{1_m}$ | $M_{2_m}$ | $f_w$ |
|---|---|---|---|---|---|---|---|---|---|
| 4.0.1 | 100 | 8 | $5 \times 10^9$ | 20 | 3.0 | 0.003 | 0.50 | 0.35 | $1.7 \times 10^{-2}$ |
| 4.0.2 | 100 | 8 | $5 \times 10^9$ | 20 | 3.0 | 0.003 | 0.85 | 0.50 | $1.5 \times 10^{-3}$ |
| 4.1.1 | 100 | 8 | $5 \times 10^9$ | 20 | 10.0 | 0.01 | 0.50 | 0.35 | $1.7 \times 10^{-2}$ |
| 4.1.2 | 146 | 10 | $5 \times 10^9$ | 20 | 10.0 | 0.01 | 0.85 | 0.50 | $1.5 \times 10^{-3}$ |
| 4.1.3 | 200 | 10 | $5 \times 10^9$ | 30 | 5.0 | 0.005 | 0.63 | 0.39 | $8.3 \times 10^{-3}$ |
| 4.2.1 | 807 | 10 | $5 \times 10^9$ | 10 | 5.0 | 0.05 | 0.65 | 0.50 | $5.9 \times 10^{-3}$ |
| 4.3.1 | 100 | $8^\star$ | $5 \times 10^9$ | 20 | 5.0 | 0.005 | 0.50 | 0.35 | $4.6 \times 10^{-2}$ |
| 4.3.2 | 100 | $8^\star$ | $5 \times 10^9$ | 20 | 5.0 | 0.005 | 0.85 | 0.50 | $6.4 \times 10^{-3}$ |
| 4.3.3 | 200 | $8^\star$ | $5 \times 10^9$ | 20 | 5.0 | 0.005 | 0.50 | 0.35 | $4.6 \times 10^{-2}$ |
| 5.0.1 | 100 | 10 | $5 \times 10^8$ | 30 | 0.5 | 0.0005 | 0.80 | 0.60 | $7.2 \times 10^{-4}$ |
| 5.0.2 | 100 | 10 | $5 \times 10^8$ | 30 | 1.0 | 0.001 | 0.50 | 0.35 | $1.7 \times 10^{-2}$ |
| 5.0.3 | 125 | 8 | $5 \times 10^8$ | 20 | 10.0 | 0.01 | 0.50 | 0.25 | $3.2 \times 10^{-2}$ |
| 5.0.4 | 100 | 10 | $5 \times 10^8$ | 100 | 0.5 | 0.05 | 0.80 | 0.60 | $7.2 \times 10^{-4}$ |

$^\star$ For Model 5 only 8 mass groups were used

**Table 4a.**

Encounter statistics

| run | $n_I$ | $n_X$ | $n_c$ | $n_{NS}$ | $n_{WD}$ | $n_{MS}$ |
|---|---|---|---|---|---|---|
| 1.0.1 | 54 | 0 | 0 | $--$ | $--$ | $--$ |
| 1.0.2 | 7 | 3 | 0 | $--$ | $--$ | $--$ |
| 2.0.1 | 0 | 0 | 0 | $--$ | $--$ | $--$ |
| 2.0.2 | 0 | 0 | 0 | $--$ | $--$ | $--$ |
| 2.1.1 | 1 | 0 | 0 | $--$ | $--$ | $--$ |
| 2.1.2 | 263 | 25 | 0 | $--$ | $--$ | $--$ |
| 3.0.1 | 4 | 6 | 1 | 0 | 0 | 1 |
| 3.1.1 | 4 | 8 | 0 | $--$ | $--$ | $--$ |
| 3.1.2 | 7 | 6 | 0 | $--$ | $--$ | $--$ |
| 3.2.1 | 4 | 2 | 4 | 0 | 2 | 2 |
| 3.2.2 | 2 | 4 | 1 | 0 | 0 | 1 |
| 3.2.3 | 2 | 9 | 0 | $--$ | $--$ | $--$ |
| 3.2.4 | 38 | 36 | 4 | 1 | 2 | 1 |
| 3.2.5 | 7 | 4 | 0 | $--$ | $--$ | $--$ |
| 3.2.6 | 13 | 13 | 2 | 0 | 2 | 0 |
| 3.3.1 | 3 | 3 | 0 | $--$ | $--$ | $--$ |
| 3.3.2 | 0 | 0 | 0 | $--$ | $--$ | $--$ |
| 3.3.3 | 2 | 0 | 1 | 0 | 0 | 1 |
| 3.3.4 | 0 | 0 | 2 | 0 | 1 | 1 |
| 3.4.1 | 27 | 7 | 1 | 0 | 0 | 1 |

**Table 4b.**

Encounter statistics

| run | $n_I$ | $n_X$ | $n_c$ | $n_{NS}$ | $n_{WD}$ | $n_{MS}$ |
|---|---|---|---|---|---|---|
| 4.0.1 | 1 | 2 | 1 | 0 | 0 | 1 |
| 4.0.2 | 1 | 1 | 0 | −− | −− | −− |
| 4.1.1 | 3 | 7 | 1 | 0 | 1 | 0 |
| 4.1.2 | 20 | 33 | 3 | 2 | 1 | 0 |
| 4.1.3 | 30 | 23 | 4 | 2 | 1 | 1 |
| 4.2.1 | 58 | 137 | 47 | 8 | 27 | 12 |
| 4.3.1 | 13 | 20 | 8 | 4 | 3 | 1 |
| 4.3.2 | 9 | 35 | 6 | 1 | 5 | 0 |
| 4.3.3 | 29 | 69 | 15 | 4 | 6 | 5 |
| 5.0.1 | 3 | 28 | 16 | 6 | 10 | 0 |
| 5.0.2 | 9 | 25 | 14 | 5 | 4 | 5 |
| 5.0.3 | 1 | 5 | 0 | −− | −− | −− |
| 5.0.4 | 6 | 31 | 4 | 2 | 2 | 0 |

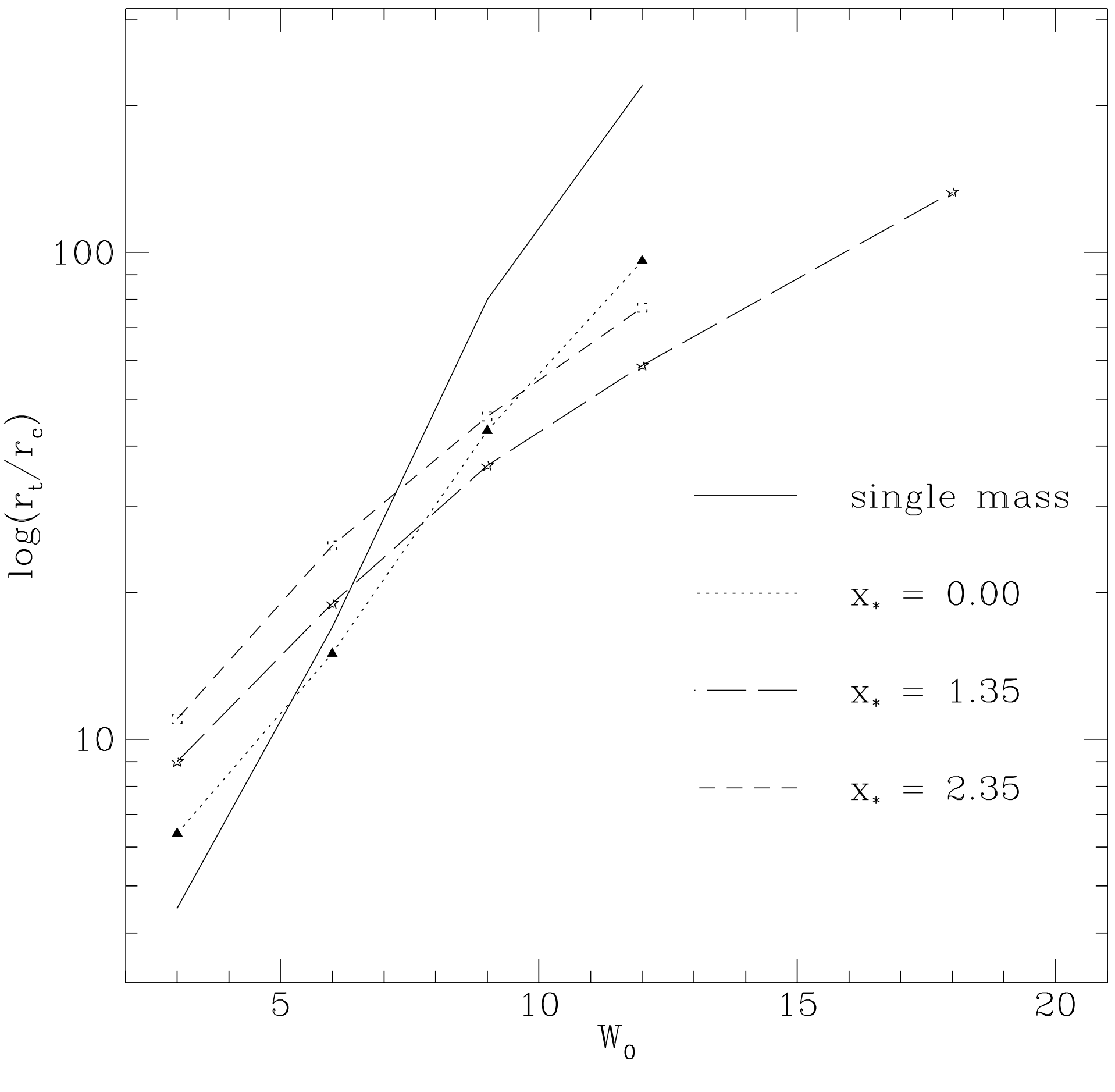

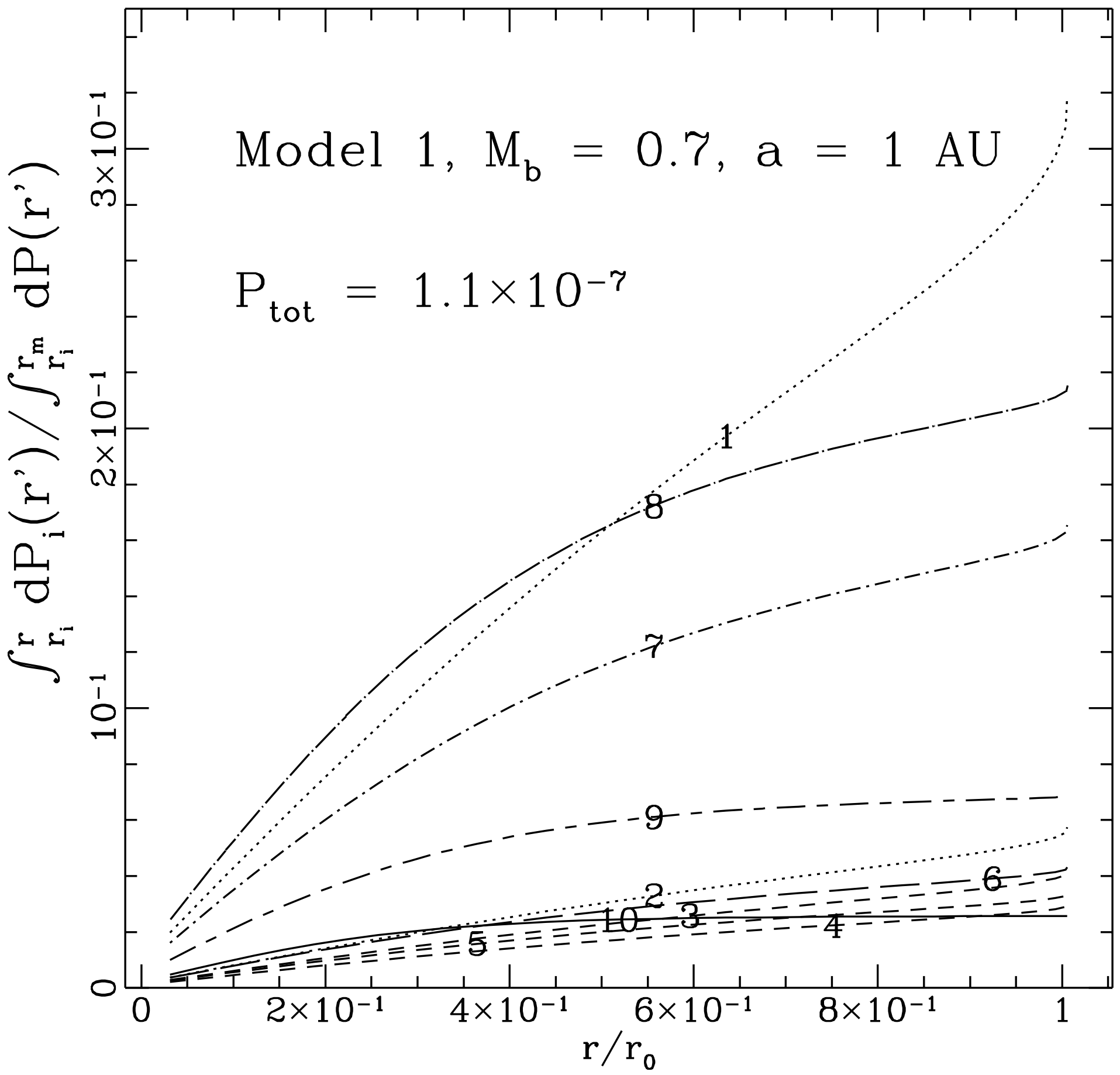

Model 1, $M_b$ = 0.7, a = 1 AU
$P_{tot}$ = $1.1\times10^{-7}$
$\int_{r_i}^{r} dP_i(r')/\int_{r_i}^{r_m} dP(r')$
$r/r_0$
0
$10^{-1}$
$2\times10^{-1}$
$3\times10^{-1}$
$4\times10^{-1}$
$6\times10^{-1}$
$8\times10^{-1}$
1
1
2
3
4
5
6
7
8
9
10

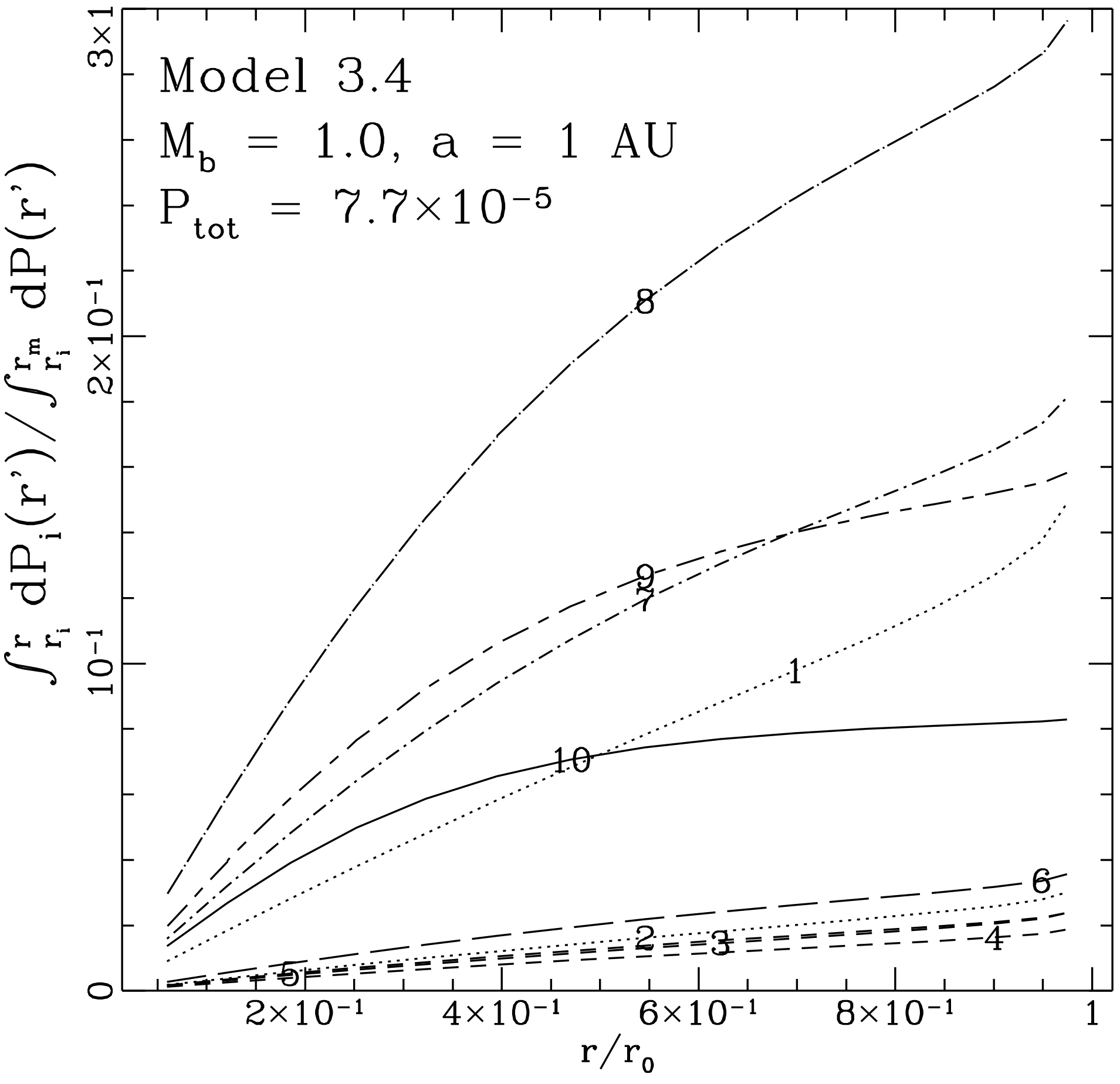

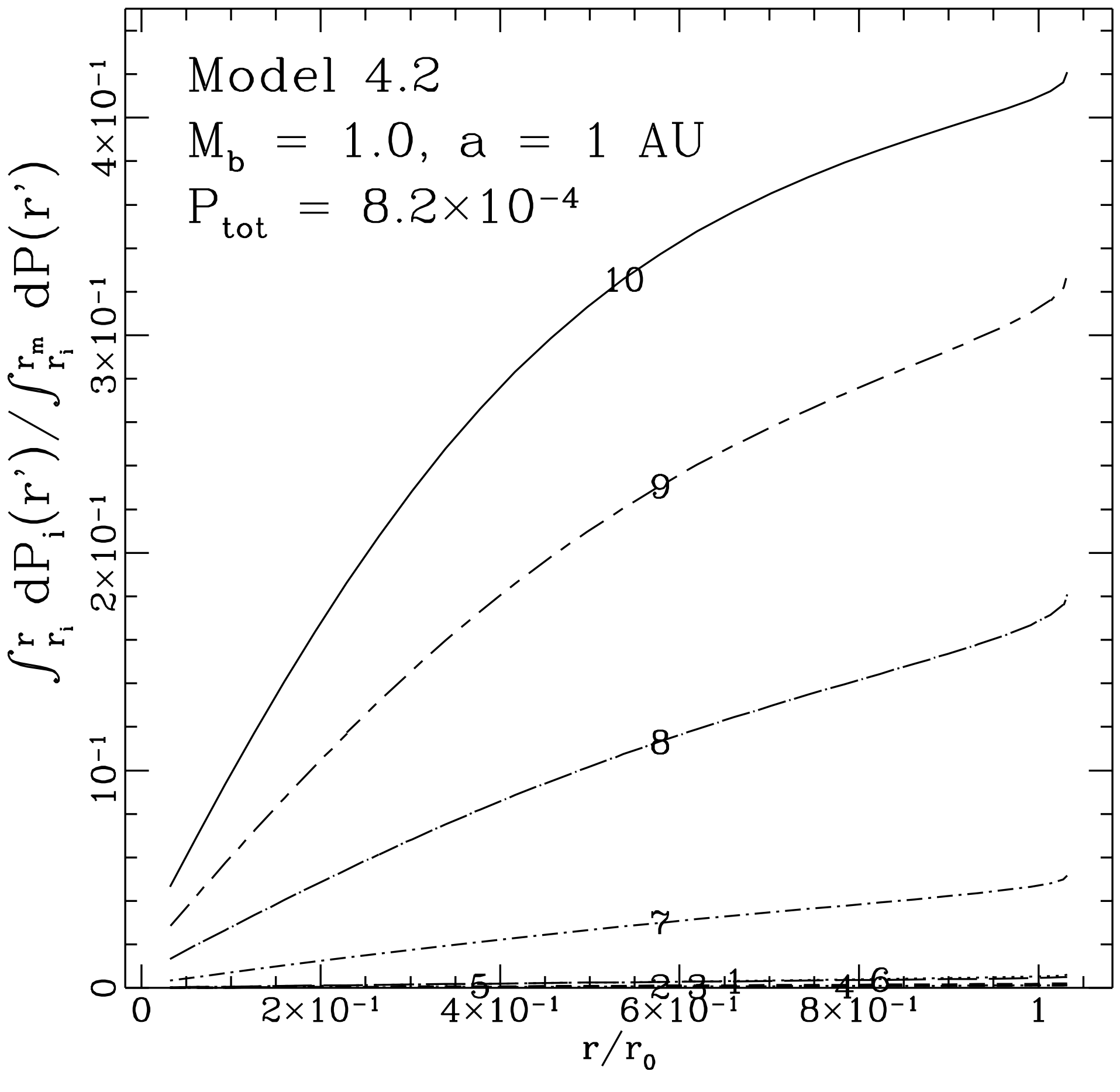

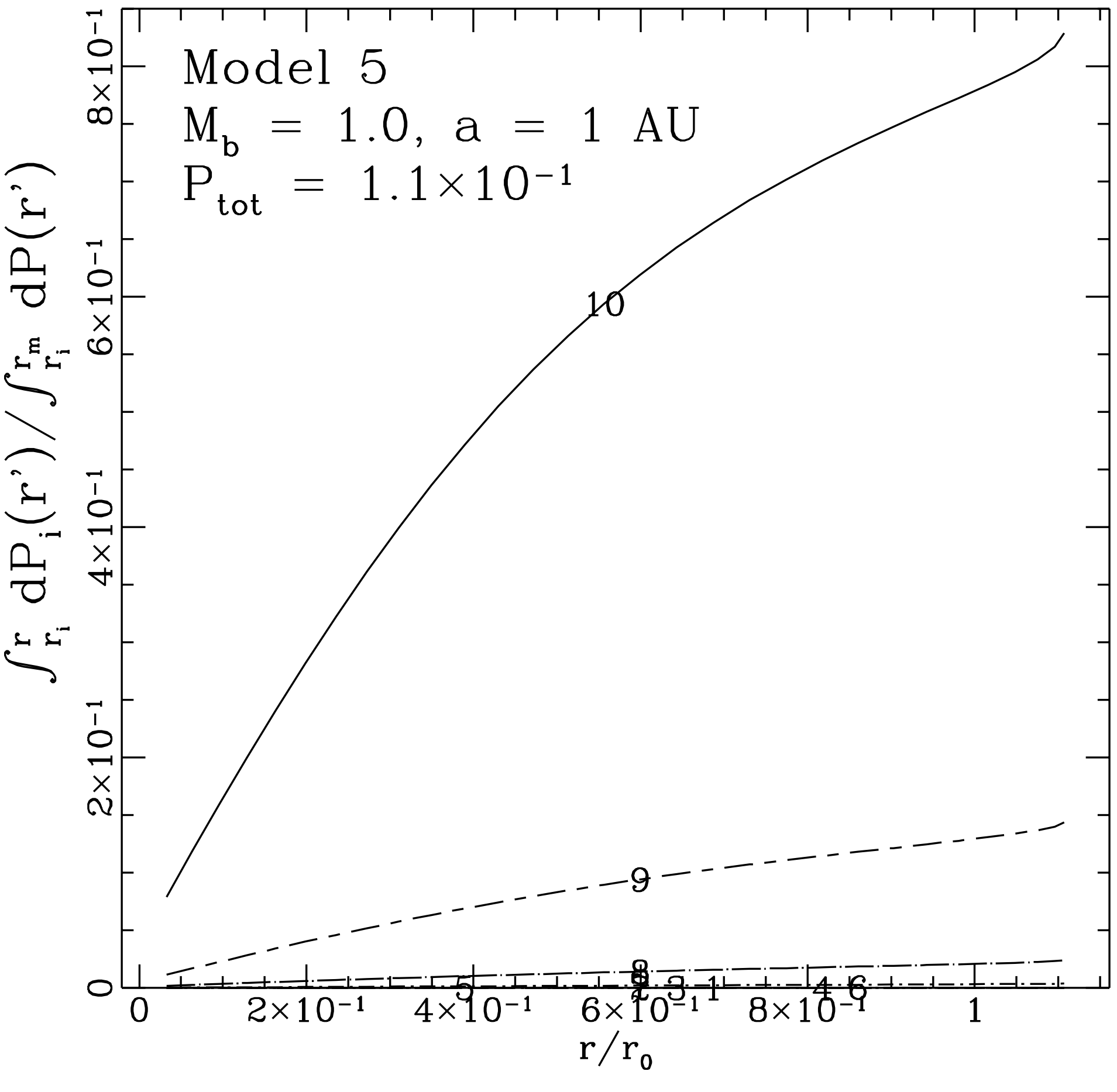

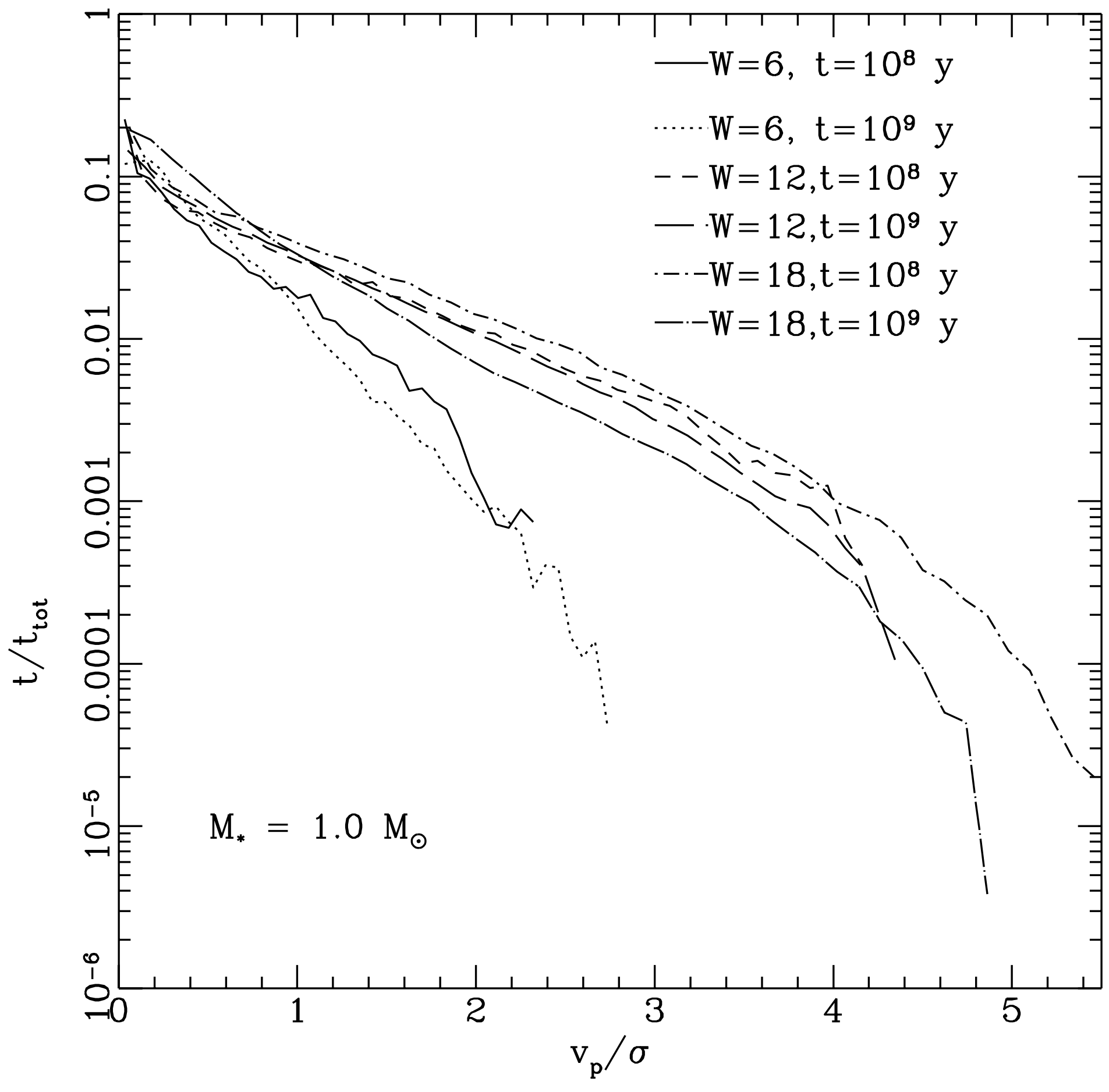

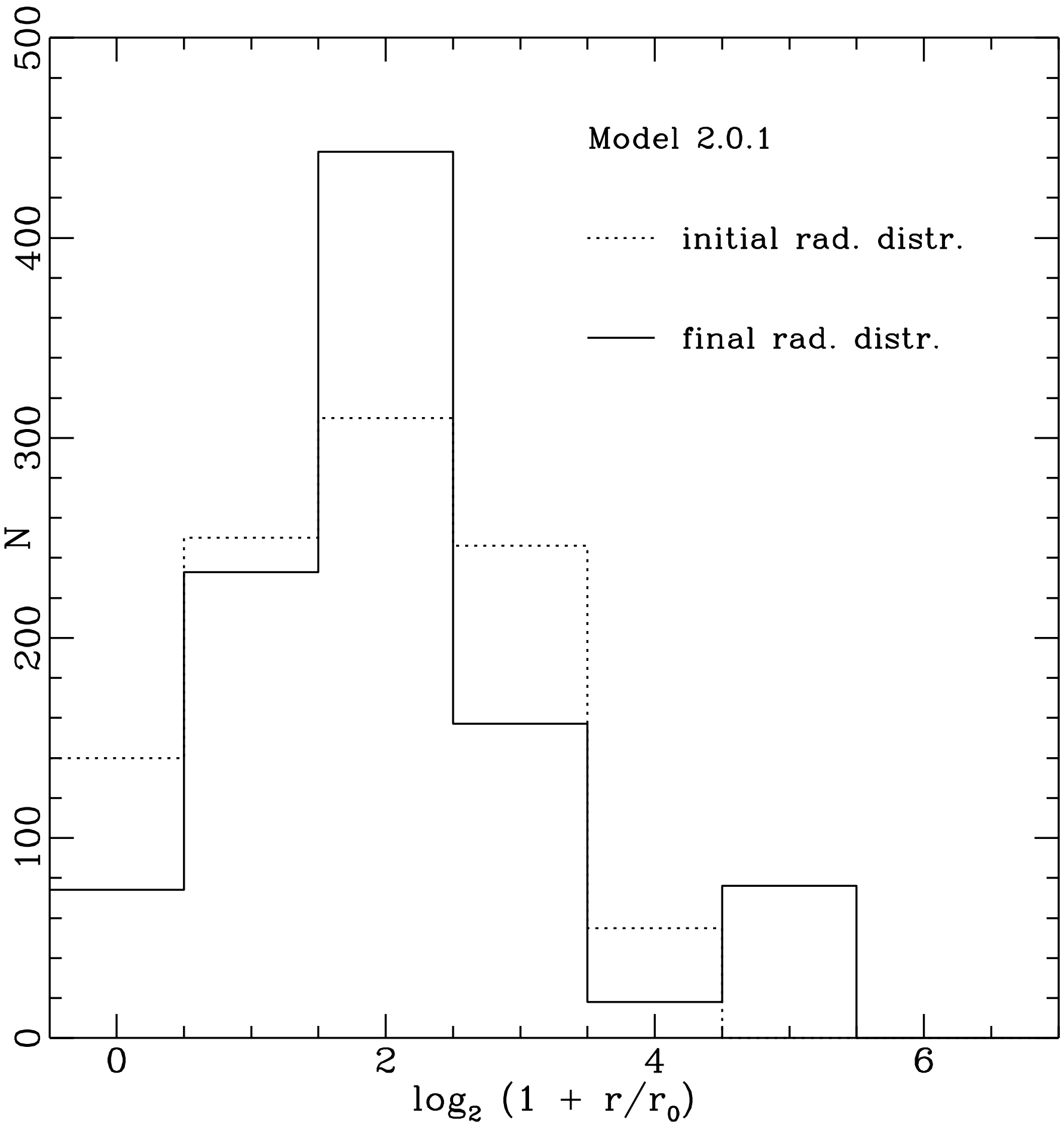

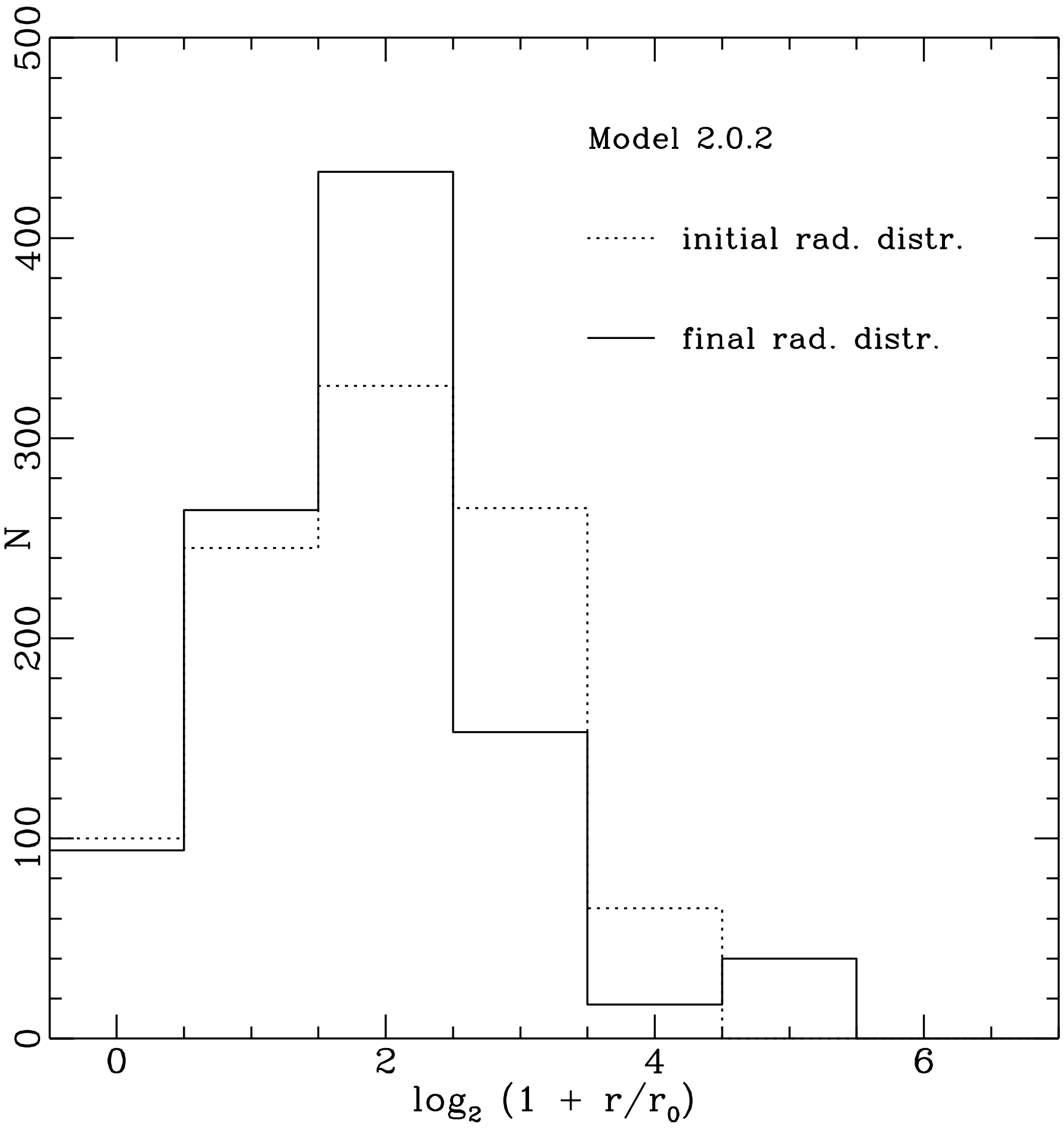

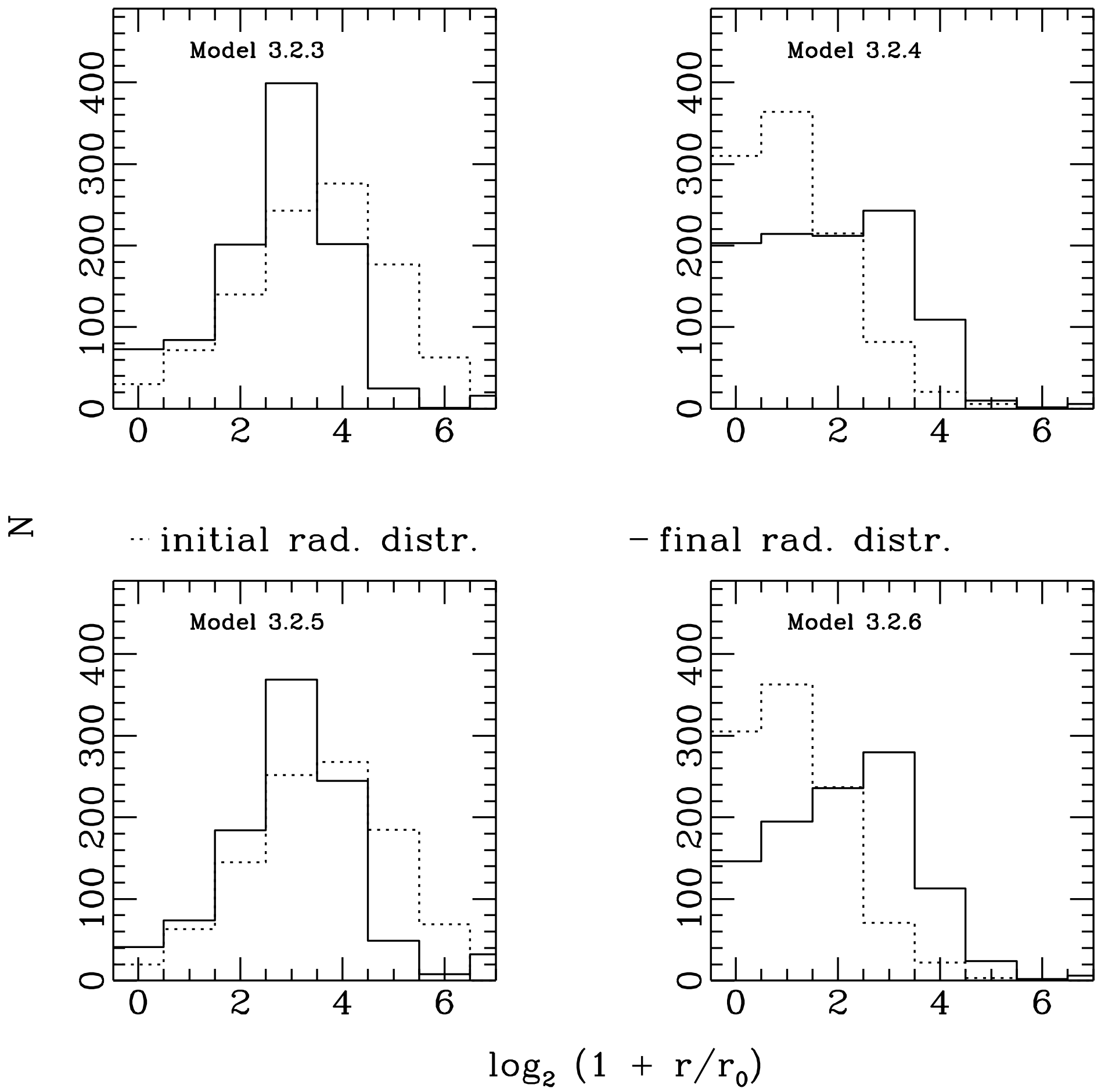

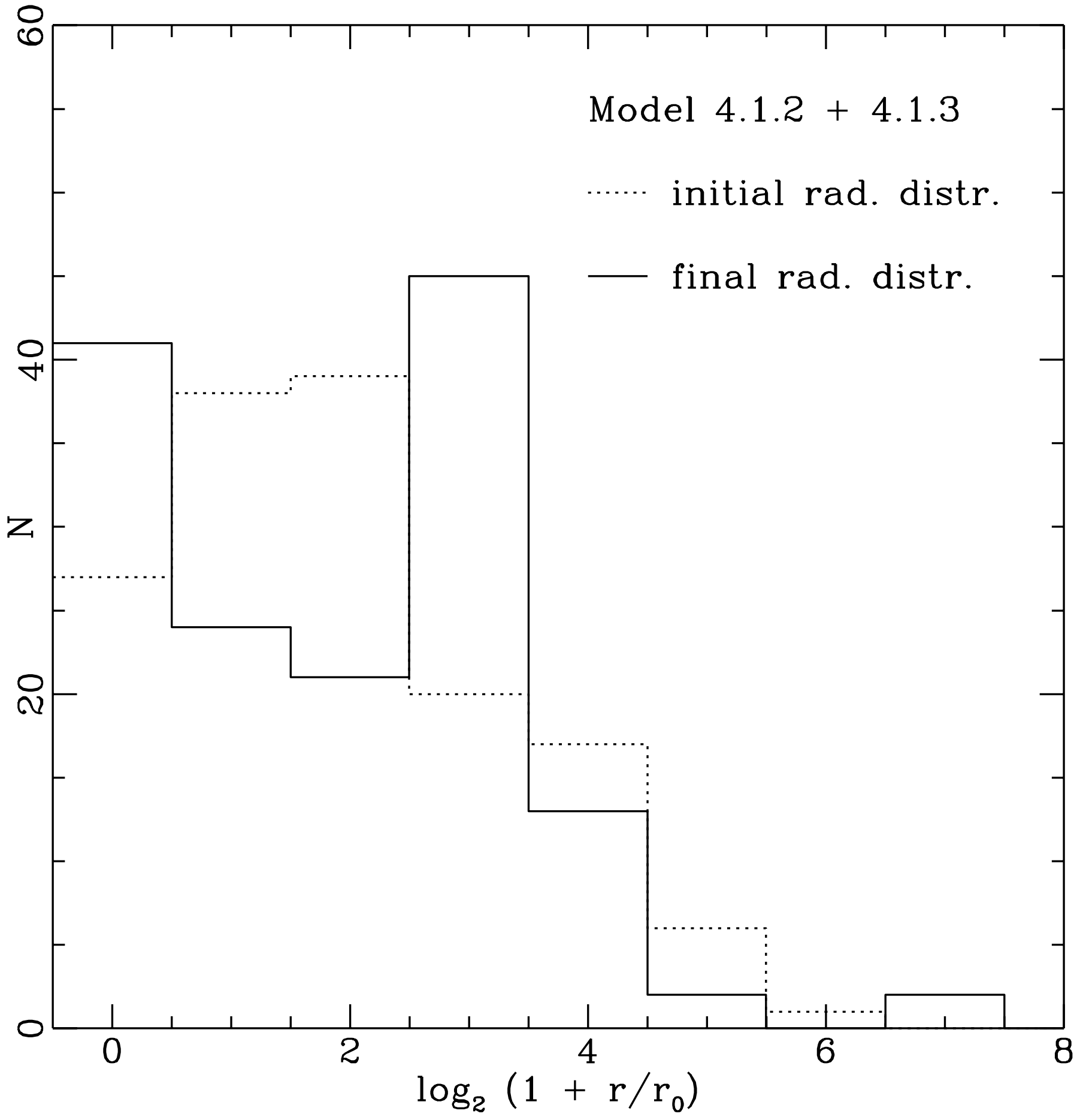

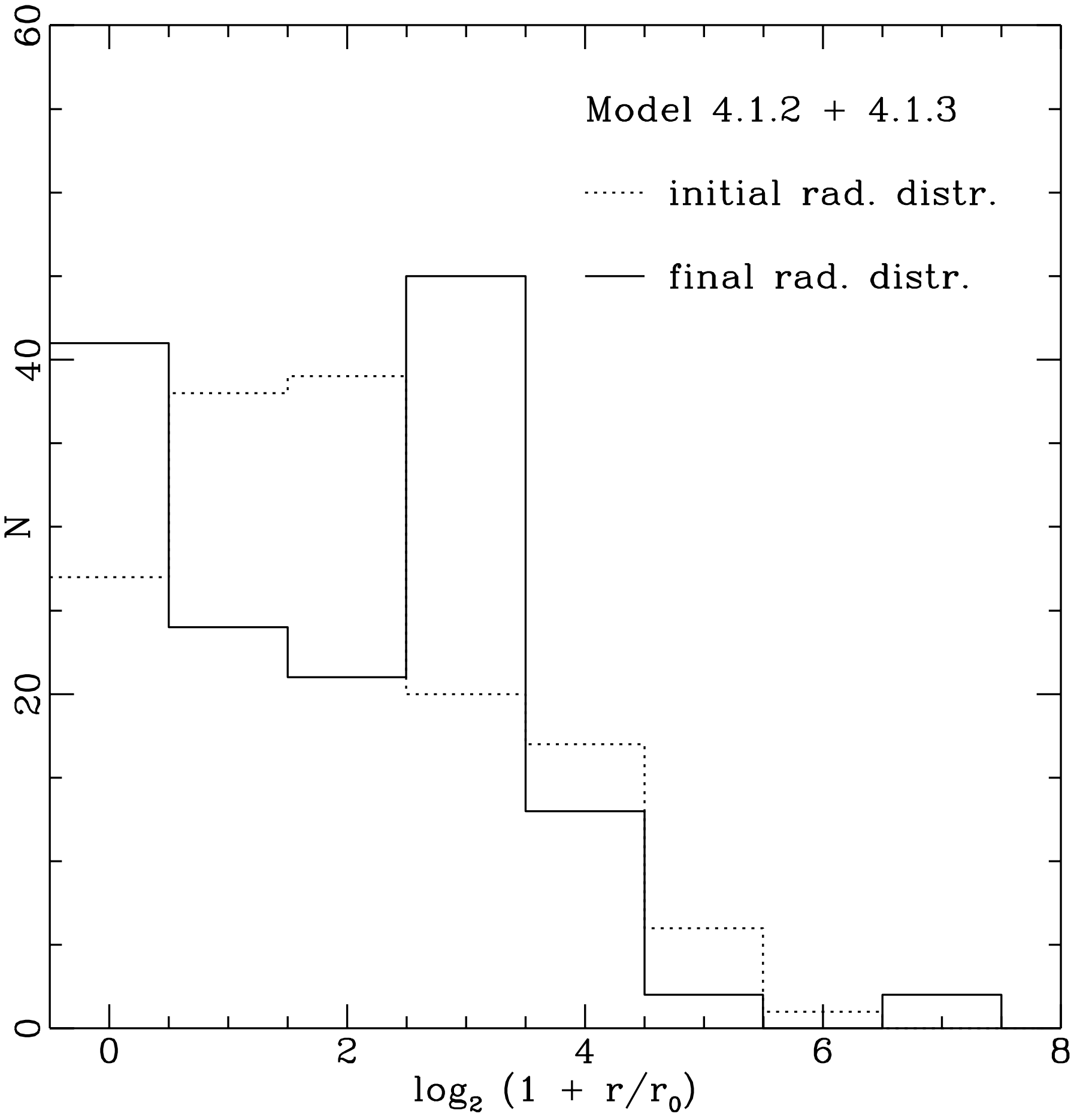

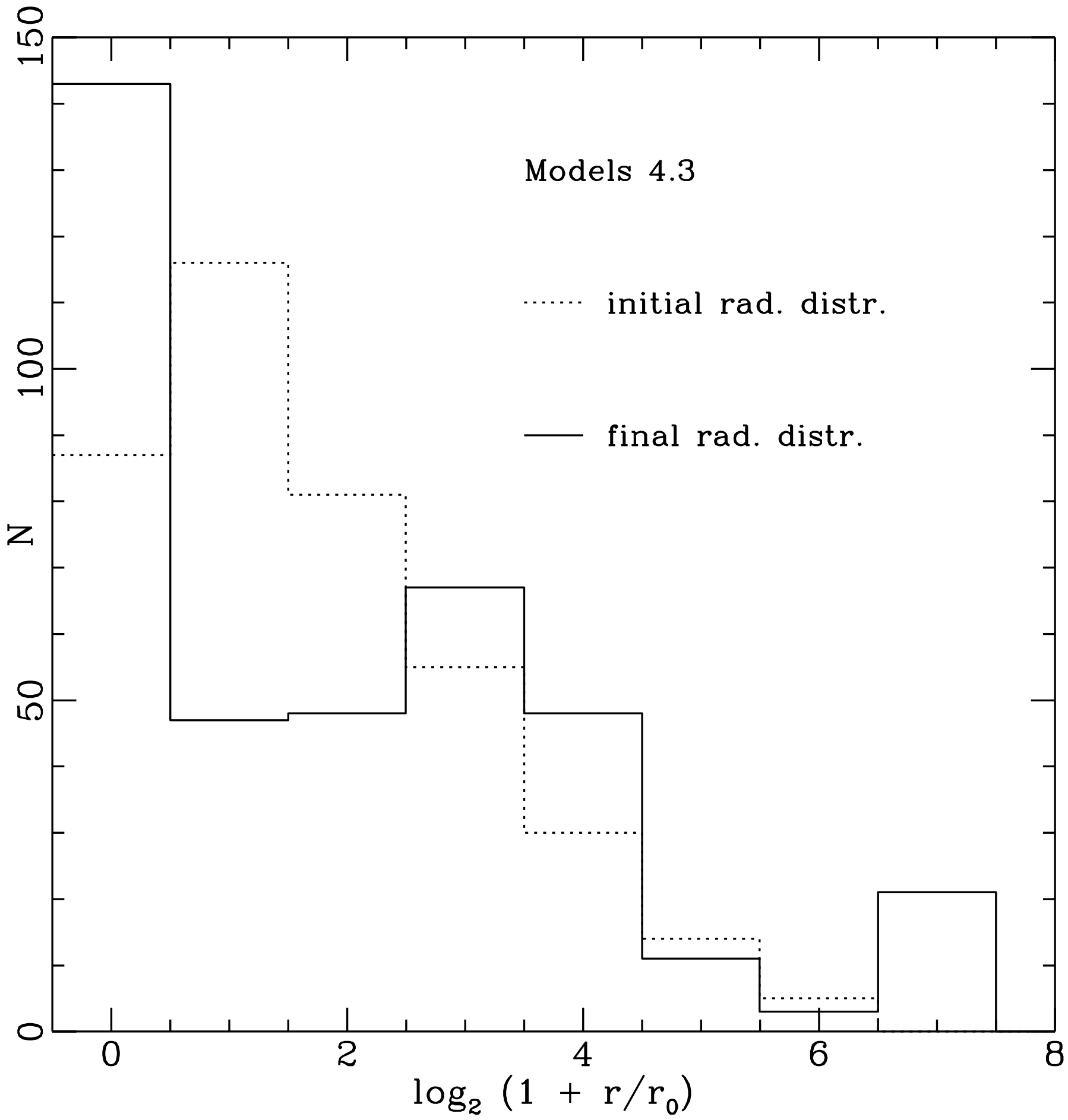

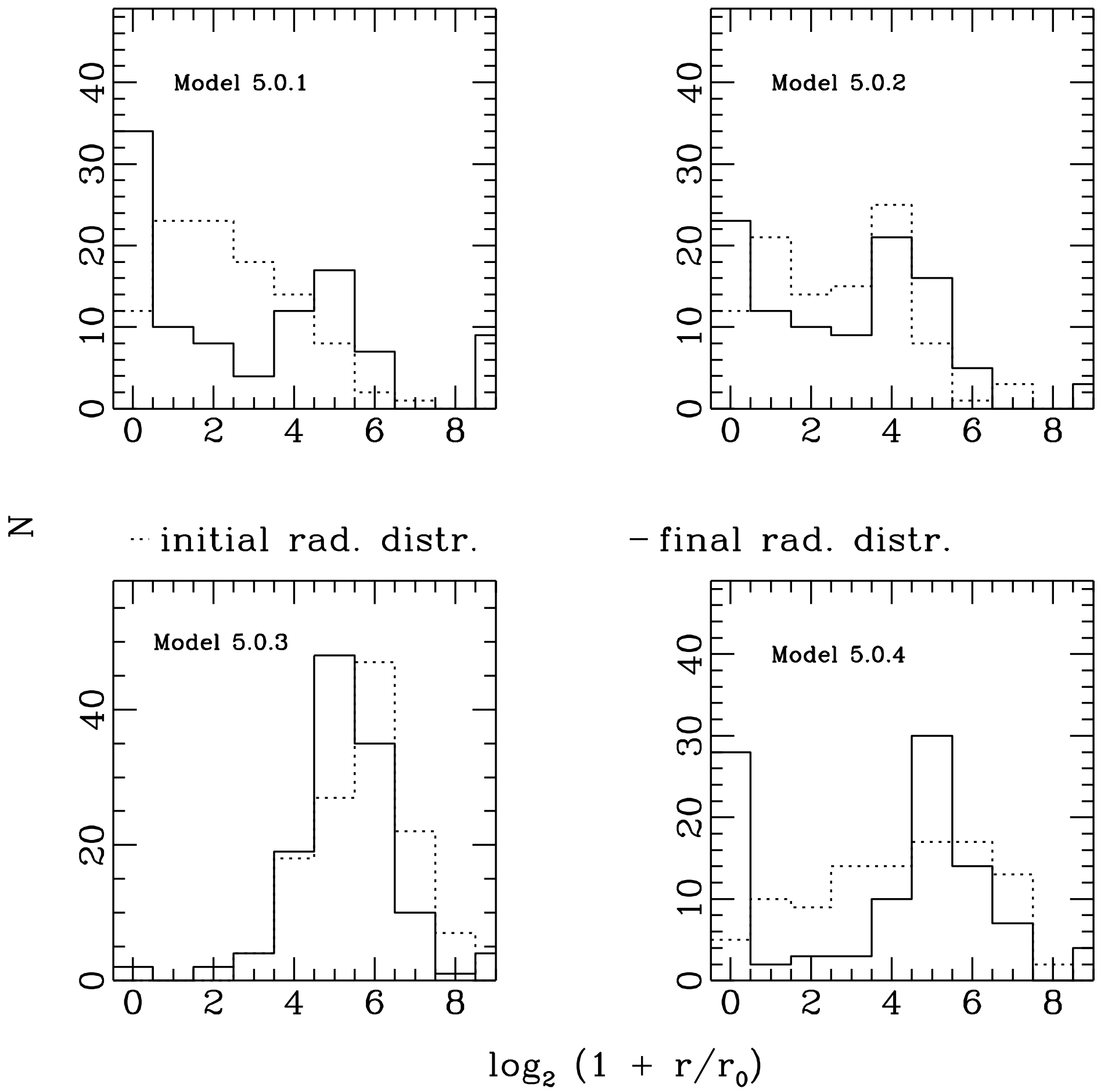

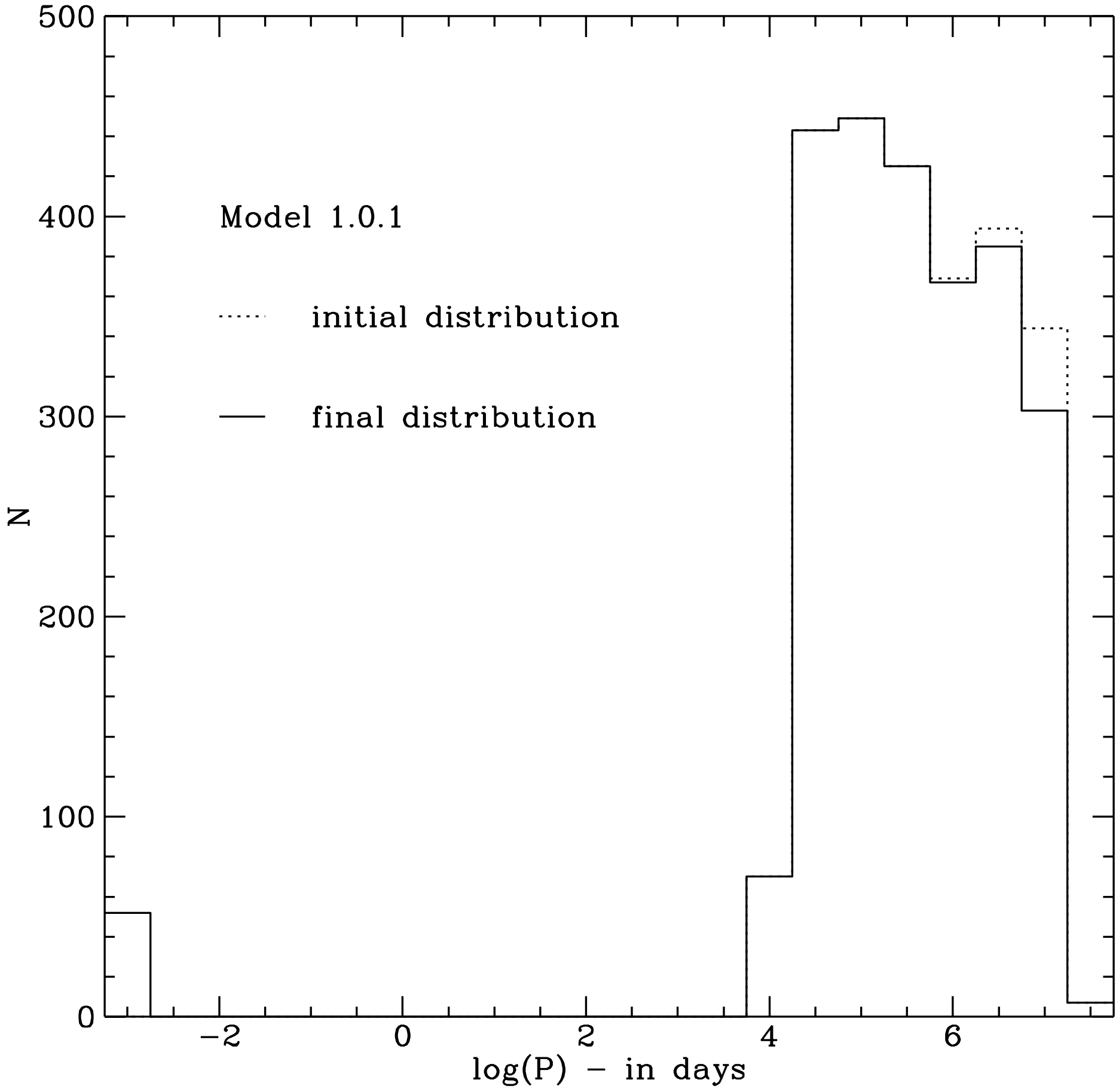

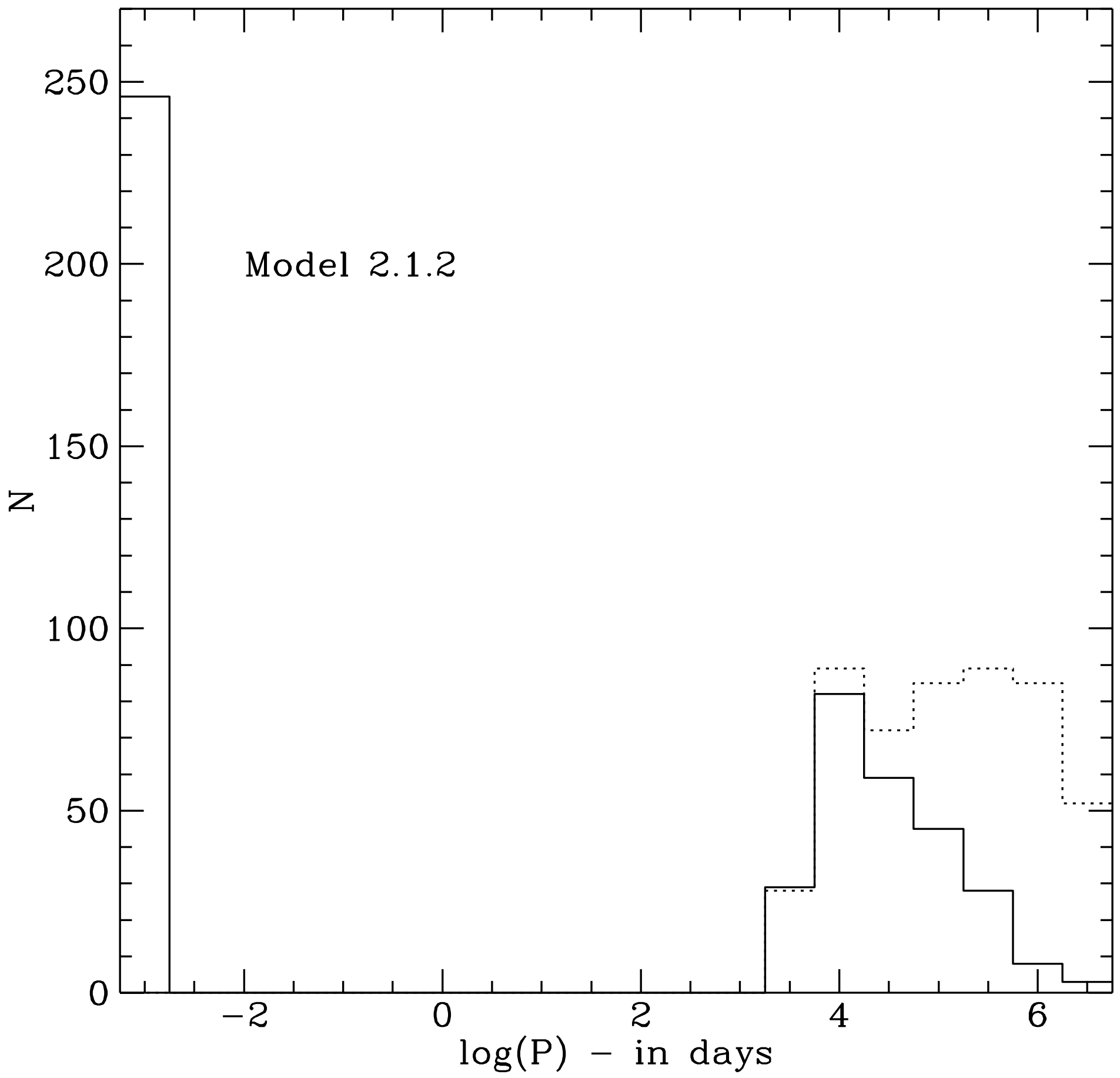

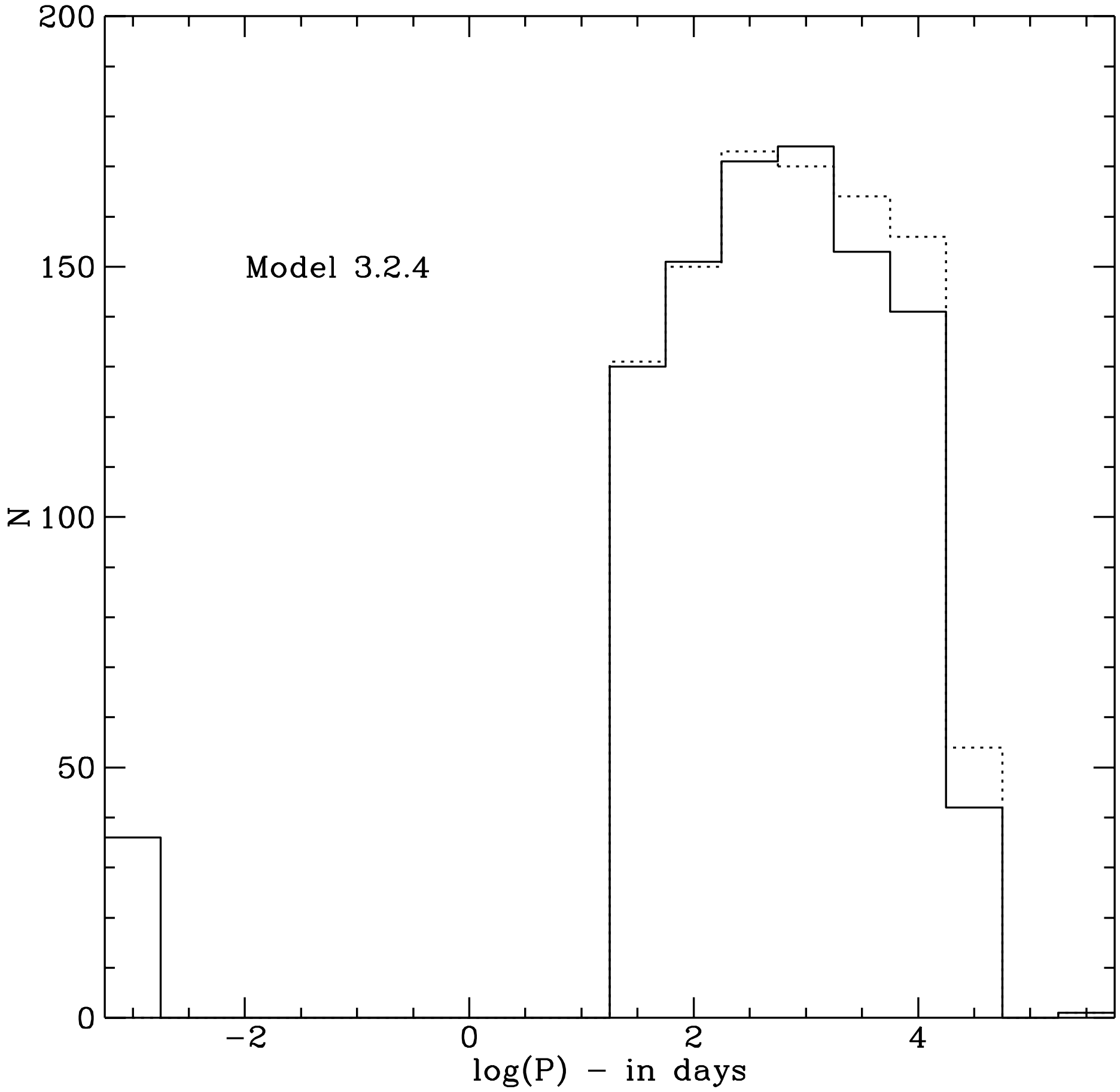

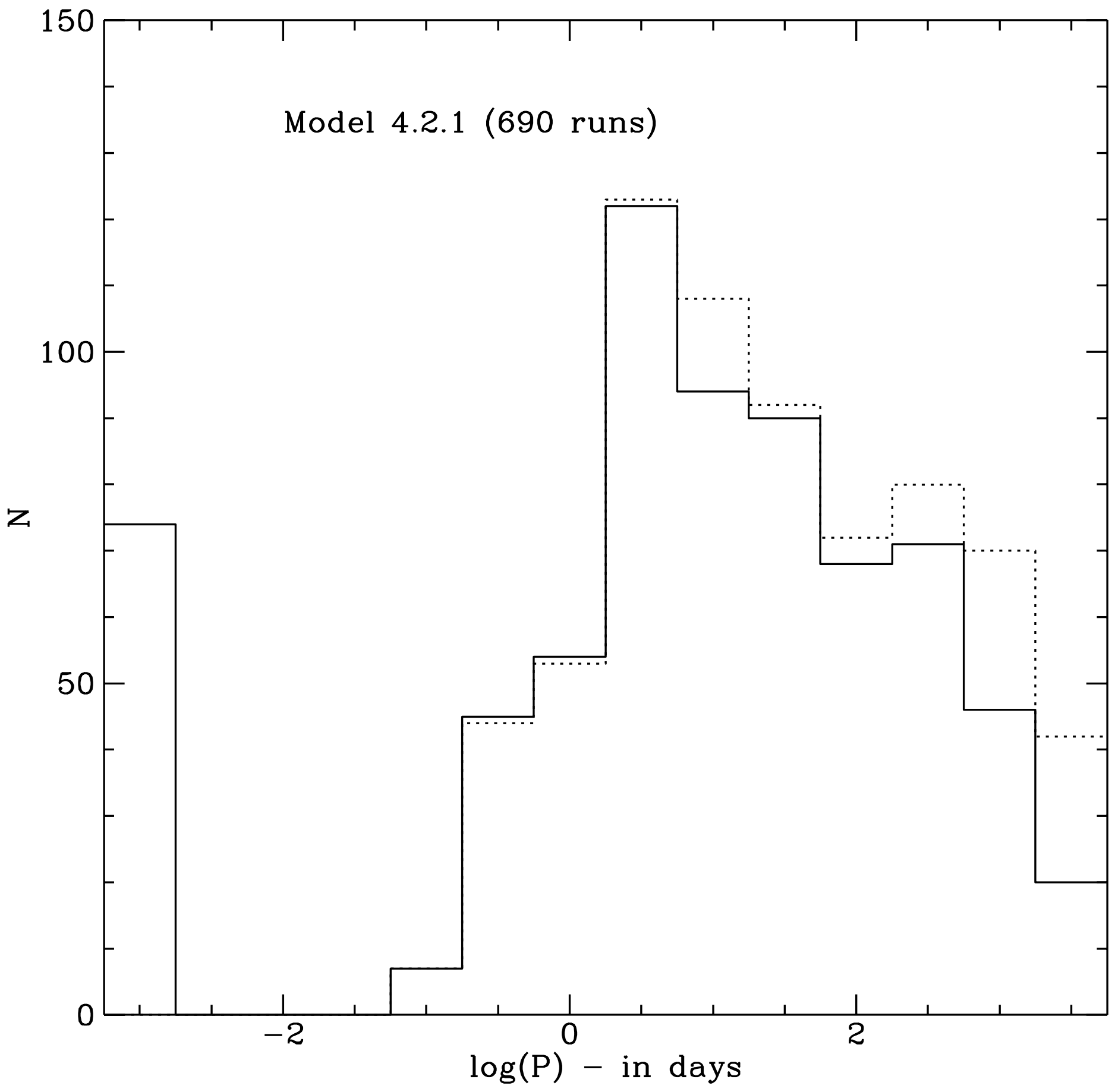

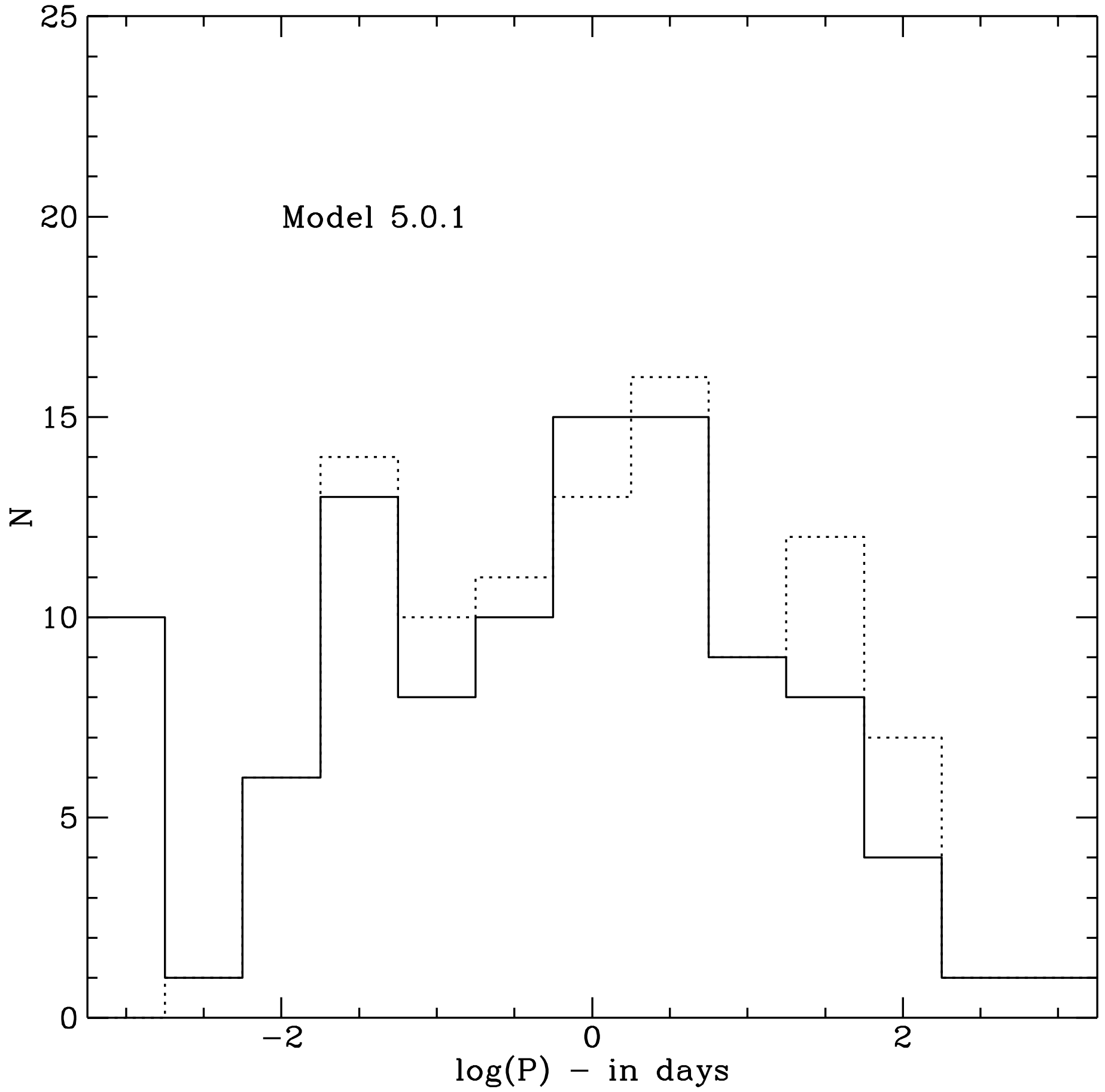